\documentclass[11pt]{article}
 \usepackage{amsmath}
\usepackage{amssymb}
\usepackage{amsthm}
\usepackage{array}
\usepackage{xy}
\usepackage{psfig}
\usepackage{subfigure}
\begin{document}

\begin{center}
\Large
{\bf The existence and detection of optically dark galaxies by 21cm surveys} \\
\normalsize
{\bf J. I. Davies, M. J. Disney, R. F. Minchin, R. Auld and R. Smith} \\
{\bf \it School of Physics and Astronomy, Cardiff University, The Parade, Cardiff, CF24 3YB, UK. \\ }
\end{center}

{\bf Abstract \\}
One explanation for the disparity between Cold Dark Matter (CDM) predictions of galaxy numbers and observations could be that there are numerous dark galaxies in the Universe. These galaxies may still contain baryons, but no stars, and may be detectable in the 21cm line of atomic hydrogen. The results of surveys for such objects, and simulations that do/do not predict their existence, are controversial. In this paper we use an analytical model of galaxy formation, consistent with CDM, to firstly show that dark galaxies are certainly a prediction of the model. Secondly, we show that objects like VIRGOHI21, a dark galaxy candidate recently discovered by us, while rare are predicted by the model. Thirdly, we show that previous 'blind' HI surveys have placed few constraints on the existence of dark galaxies. This is because they have either lacked the sensitivity and/or velocity resolution or have not had the required detailed optical follow up. We look forward to new 21cm blind surveys (ALFALFA and AGES) using the Arecibo multi-beam instrument which should find large numbers of dark galaxies if they exist. \\
{\bf Keywords \\}
Galaxies: Galaxy formation, Dark matter

\section{Introduction}
The currently favoured Cold Dark Matter (CDM) model of galaxy formation predicts many more Dark Matter (DM) halos than are observable as galaxies. Amongst other speculations the existence of dark galaxies offers one solution (Verde et al., 2002). It is possible that the currently observable galaxies are just some fraction of a larger and more varied population of DM halos. So, in addition to the known galaxies there maybe DM halos with no baryons and/or those with baryons, but no stars (Hawkins, 1997, Jimenez et al., 1997). Without star formation the latter dark galaxies should remain un-enriched in heavy elements and contain proportions of ionised, atomic and molecular hydrogen commensurate with their gas densities and temperatures. Given that these dark galaxies contain baryons they may be accessible to observation by methods tried and tested on their more visible cousins.

The exsistence of objects that contain HI, but no stars, has been known for many years. For example High Velocity Clouds, tidal tails and clouds close to optically bright galaxies, but none of these have the characteristics that might lead them to be described as galaxies. To count as a dark galaxy a source requires HI emission over galactic scales (greater than a few kpc) with velocity widths and profiles commensurate with a dynamically stable system (velocities greater than a few tens of km s$^{-1}$). 

The idea that there are galaxies only detectable by their gas content is also not new and there have been in the past blind HI surveys that might have discovered such objects if indeed they exist. Henning (1995) carried out a blind HI survey using the NRAO 300ft telescope covering 183 sq deg of sky. Most of the survey area was within the Zone of Avoidence (ZOA) thus making a convincing HI detection of a dark galaxy difficult. The Arecibo HI Sky Survey (AHISS, Zwaan et al. 1997, Zwaan 2000) covered just 13 sq deg and it appears that each detection has an associated optical counterpart. The Arecibo Slice Survey (ASS, Schneider et al. 1998, Spitzak and Schneider 1998) covered 55 sq deg. All but one of the detected galaxies has been associated with an optical source. The Arecibo Dual-Beam Survey (ADBS, Rosenberg and Schneider 2000) covered an area of 430 sq deg producing 11 objects that had no optical counterparts on the POSSI/II photographic plates. The HI Jodrell All Sky Survey (HIJASS, Lang et al. 2003) has covered about 1100 sq deg, but at the moment lacks the optical follow up to decide on the dark galaxy issue. These surveys covered small areas compared to the recent HIPASS observations (the catalogue of observations is known as HICAT, Meyer et al. 2004). The HIPASS has now covered the southern sky, $\approx$ 21,000 sq deg, using a multi-beam instrument (HIDEEP is a smaller deeper sub-set of HIPASS). One apparent result to come from this work is that all HI sources seem to be associated with an optical counterpart (Doyle et al. 2005) and on the face of it there seems to be little scope for the existence of dark galaxies. This has important implications: for example with regard to the efficiency with which gas is constrained to remain within DM haloes and the close physical link between the availability of HI and the formation of stars.  In table 1 we list the areas covered $(A)$, channel velocity widths ($\delta V$) and rms noise ($\sigma_{rms}$) of the above surveys. We will return to what each might and might not have expected to detect later. In table 1 we have also included the Arecibo Legacy Fast ALFA survey (ALFALFA survey - Giovanelli et al., 2005) which is currently being carried out at the Arecibo observatory and the Arecibo Galactic Environments Survey (AGES - www.astro.cardiff.ac.uk/groups/galaxies/ages.html) which is due to start in January 2006.

\begin{table}
\begin{center}
\begin{tabular}{l|c|c|c|c}
Survey  & $A$        &  $\delta V$    & $\sigma_{rms}$    &  $V^{Max}$ \\
        & (sq deg) & (km s$^{-1}$)  & (mJy beam$^{-1}$) &   (km s$^{-1}$)  \\ \hline
Henning &    183   &      22        &     3.4           &  6800 \\
AHISS   &  13      &      16        &     0.8           &  7400  \\
ASS     &  34      &      16        &     2.0           &  8340   \\
ADBS    &  430     &      34        &     3.5           &  7977   \\
HIJASS  &  1,115   &      13        &     14.0          &  10000  \\
HIPASS  &  21,000  &      13        &     13.0          &  12700  \\
HIDEEP  &  32      &      13        &     3.4           &  12700  \\
VIRGOHI &  32      &      13        &     4.0           &  10000   \\
ALFALFA &  10,000  &      5         &     2.0           &  18000  \\
AGES    &  200     &      5         &     0.5           &  18000 \\
\end{tabular}
\end{center}
\caption{A comparison of survey area (A), velocity channel widths ($\delta V$), rms noise ($\sigma_{rms}$) and maximum velocity ($V^{Max}$) for previous and future  (ALFALFA and AGES) blind HI surveys.}
\end{table}

In 2004 we used the Jodrell Bank multi-beam instrument to carry out a deep survey of part of the Virgo cluster (Davies et al., 2004). The Virgo survey (VIRGOHI) reduced the $\approx 14$ mJy per channel of the HIJASS survey to $\approx 4$ mJy (Table 1) enabling us to look for both lower mass and lower column density hydrogen than would be detected in the normal survey. The area observed was also one in which we know there is a large number of nearby optically detected galaxies. Davies et al. identified four objects that had HI detections but no obvious optical counterparts. One of these was subsequently rejected as noise, a second was later identified, after obtaining new deep CCD data, with a very faint dwarf galaxy. The third object has subsequently been associated with debris stripped from NGC4388 (Oosterloo and van Gorkom, 2005). The fourth object, VIRGOHI21, remains as a strong candidate for a dark galaxy (Minchin et al. 2005, Minchin et al. 2006). VIRGOHI21 is extended over a region of about 10 kpc and has a HI velocity width of about 200 km s$^{-1}$ - according to the HI data it is a typical galaxy, but there is no optical counterpart down to very low surface brightness limits.

 In this paper we first of all want to re-address the issue of whether dark galaxies can exist within our current models of galaxy formation and particularly examine the recent contention by Taylor and Webster (2005) that almost all galaxies containing HI will form stars. Secondly we want to see if a dark galaxy with the characteristics of VIRGOHI21 is a possiblity. Thirdly we want to discuss whether the existence of dark galaxies is compatable with previous blind HI surveys. Verde et al. (2002) have also produced models very similar to those discussed in this paper and predict the existence of dark galaxies, we concentrate on the detectability of these galaxies by 21cm surveys.

\section{The model}
We have used the prescription of Mo et al., (1998) [MMW] to model thin self-gravitating baryonic galactic discs within NFW (Navarro, Frenk and White, 1995,1996) dark matter halos. In our view the MMW model is as good as any other and more transparent than most, in formulating the predictions of galaxy formation theory. The model is given in outline below - much fuller details are given in MMW. 

We select a DM halo mass (M) for each galaxy simulated from a Schechter function 
\begin{equation}
\Theta(M) dM = (M/M^{*})^{-\alpha} \exp{-(M/M^{*})} d(M/M^{*})
\end{equation}
with $M^{*}=10^{12}$ $M_{\odot}$, low mass slope $\alpha=-1.5$ and mass range $10^{8}-10^{13}$ $M_{\odot}$ (Dalcanton et al 1997). Each halo has a NFW profile concentration index C=10 (MMW). A fraction $m_{d}$ of the mass of each halo collapses to form a baryonic disc embedded in the halo. The fraction of the total angular momentum associated with the disc ($j_{d}$) is the same as the fraction of the mass associated with the disc ($j_{d}$=$m_{d}$) (MMW). The galaxy spin parameter ($\lambda$) is taken from the lognormal distribution
\begin{equation}
p(\lambda) d\lambda = \frac{1}{\sigma_{\lambda}\sqrt{2\pi}}\exp{\frac{ln^{2}(
\lambda/<\lambda>)}{2\sigma_{\lambda}^{2}}} \frac{d\lambda}{\lambda}
\end{equation}
where $<\lambda>=0.05$ and $\sigma_{\lambda}=0.5$ (Warren et al., 1992). As in MMW we assume that our galaxies form at $z=0$ ($H(z)=1.0$) and that $h=0.72$ ($H_{o}=72$ km s$^{-1}$ Mpc$^{-1}$). We calculate the maximum circular velocity $V_{c}$ in km s$^{-1}$ of each disc using
\begin{equation}
V_{c} = V_{200} f_{V}
\end{equation}
where
\begin{equation}
 V_{200}=0.017 (M H(z) h)^{1/3}
\end{equation}
and $f_{V}$ is a dimensionless factor obtained from the fitting function given in MMW (accurate to $\approx$ 20\%). The exponential scale length of each disc ($R_{d}$ in kpc) is calculated using
\begin{equation}
R_{d}=R_{d}^{iso} f_{c}^{-0.5} f_{R}
\end{equation}
where
\begin{equation}
R_{d}^{iso}=0.7 (H(z) h)^{-1.0} \lambda V_{200} (j_{d}/m_{d})
\end{equation}
and $f_{c}$ (accurate to $\approx$ 1\%) and $f_{R}$ (accurate to $\approx$ 20\%) are again dimensionless fitting functions given in MMW.
We can also calculate a disc central gas column density in cm$^{-2}$ (assuming at this point that we have a gas disc) using
\begin{equation}
\Sigma_{0}=\frac{2.0 \times 10^{13} m_{d} M} {R_{d}^{2}}
\end{equation}
The above model produces discs with a wide range of properties, but not all of them will be stable. Following MMW we use a disc stability criterion ($e$) to assess the stability of each disc (see Efstathiou et al. 1982). Only discs with $e>1.0$ will be included in our final list of galaxies.
\begin{equation}
e=\frac{1}{2^{1/4}}(\lambda^{\prime}/m_{d})^{0.5} f_{c}^{-1/4} f_{R}^{1/2} f_{V}
\end{equation}
where $\lambda^{\prime}=\lambda j_{d}/m_{d}$.

The above model can be used to produce a simulated sample of stable gaseous discs embedded within DM halos. For each halo we have a total mass (M), disc mass ($m_{d}M$), disc circular velocity ($V_{c}$), exponential disc scale length ($R_{d}$) and central gas column density ($\Sigma_{0}$). Our next concern is to distinguish between those stable discs that form stars and those that remain dark. 

There have previously been a large number of papers that have considered the physical conditions required for gas to turn itself into stars. Many of these suggest a surface density threshold (Spitzer 1968, Quirk 1972, Fall and Efstathiou 1980, Kennicutt 1989) and most of these are based around the Toomre criterion for a locally stable gas disc (Toomre 1964, Goldreich and Lyndon-Bell 1965, Binney and Tremaine 1987). A recent comprehensive discussion of the Toomre criterion and its applicability to the star formation process is given in Schaye 2004. Schaye also used a model very similar to the one described above to investigate the conditions under which gas can cool to form stars. He implements the photoionisation package CLOUDY to consider the influence of the gas column density, gas metalicity and background ionising radiation on the ability of the gas to cool. The result of this study is that Schaye agrees that there is a threshold surface density below which star formation is inhibated (Skillman 1987). This critical surface density is $\Sigma_{Crit} \approx 5 \times 10^{20}$ cm$^{-2}$.
\footnote{Schaye (2004) actually says that this critical density lies between 3 and 10 $\times 10^{20}$ cm$^{-2}$ and that this is '....insensitive to the exact values of model parameters such as the intensity of the UV radiation, the metalicity, the turbulent pressure, and the mass fraction in collisionless matter.'. For an example of this dependency see his equation 23. If the intensity of the UV radiation is reduced to that of the present day background (Schaye assummes some heating by stars) then the critical column density is reduced by about a factor of two. But given that we might also expect the metalicity of the gas to be very low, because there has been no star formation, one could very easily argue that the critical density we are using is too low and more galaxies remain dark. In the absence of good information either way we adopt Schaye's approximate value for the star formation threshold.} 
We thus distinguish within our simulation between stable galaxy discs that can form stars ($\Sigma_{0} > 5 \times 10^{20}$ cm$^{-2}$) and those that cannot ($\Sigma_{0} < 5 \times 10^{20}$ cm$^{-2}$).

Before discussing the properties of the simulated galaxies we have one parameter that we have not yet put a value to - $m_{d}$ (following MMW we have set $j_{d}$ = $m_{d}$). $m_{d}$ is the fraction of the halo mass that forms the disc and the assumption is that this is in the form of baryonic matter. $m_{d}$ turns out to be the crucial parameter when deciding whether dark galaxies are permissible within the above galaxy formation scenario. For a small value of $m_{d}$ (see below) many low column density discs are formed that are not able to form stars - dark galaxies (see fig 1).

\begin{figure}
\begin{center}
\psfig{figure=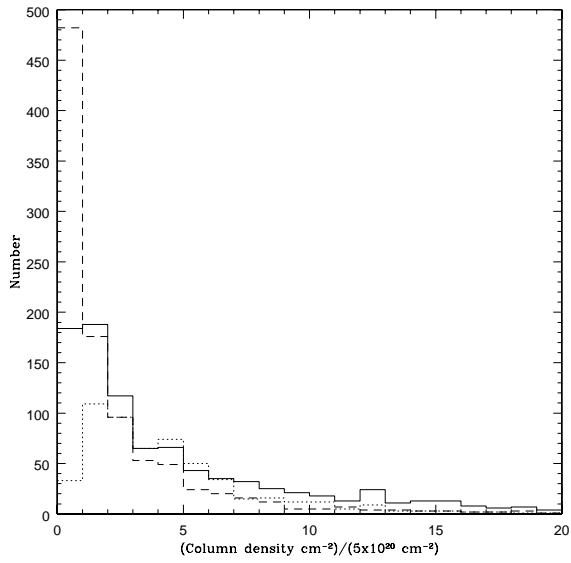,width=8cm}
\end{center}
\caption{The distribution of column densities (normalised by $\Sigma_{Crit}=5 \times 10^{20}$ cm$^{-2}$) for three values of $m_{d}$. Large dash $m_{d}=0.01$, solid line $m_{d}=0.025$, small dash $m_{d}=0.05$. Only objects in the first column on the left will remain as dark galaxies. We see that small values of $m_{d}$ lead to relatively large numbers of dark galaxies.}
\end{figure}

\begin{figure}
\Large
{\bf a)}
\normalsize
\subfigure{\psfig{figure=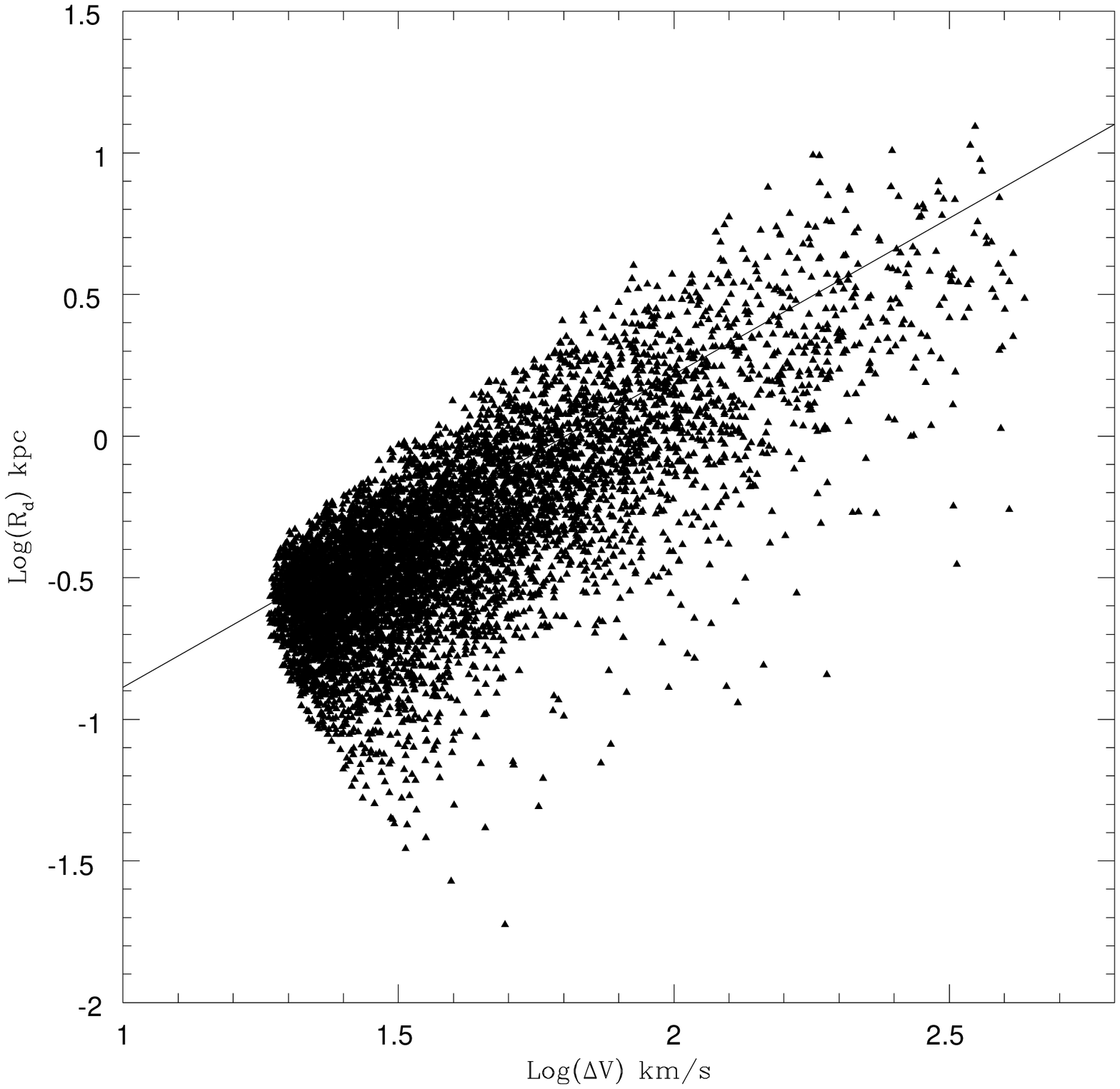,width=8cm}}
\Large
{\\ \bf b)}
\normalsize
\subfigure{\psfig{figure=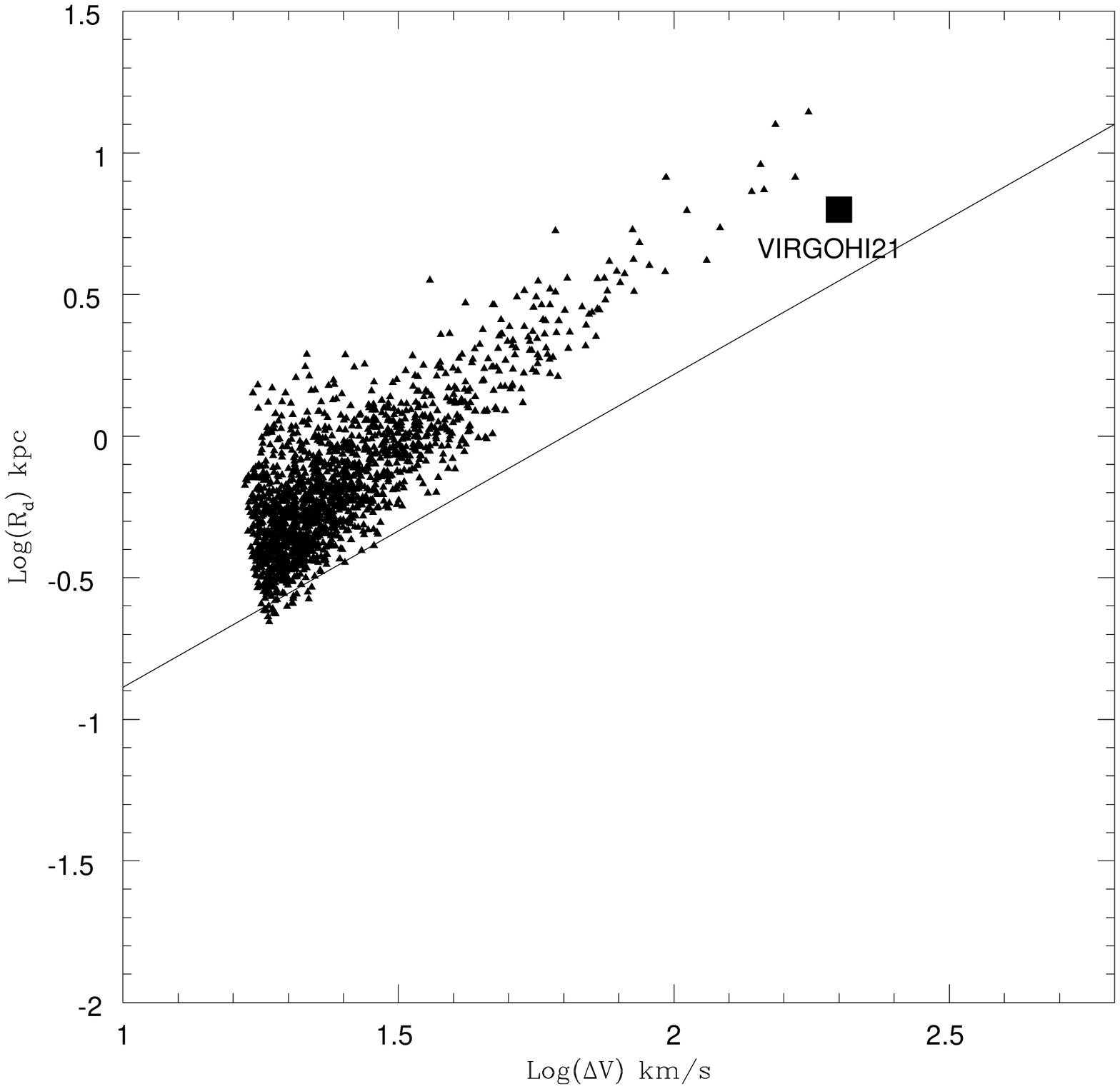,width=8cm}}
\caption{The range of galaxy velocity widths and exponential disc scale sizes for galaxies that can form stars (a) and those that cannot (b). The line corresponds to the solid line drawn on fig. 4c of MMW. This line passes roughly through the middle of the observational data in the range $2.0< \log{\Delta V}<2.5$ and $0.0<\log{R_{d}}<1$ and so our simulation compares well with the observations of large disc galaxies. As is expected from the mass function that we are using the simulation has many more small, low velocity discs compared to the observational data used by MMW. The position of VIRGOHI21, a candidate dark galaxy, is indicated on 2b.}
\end{figure}

 An upper bound on $m_{d}$ might be taken to be $\Omega_{Baryon}/\Omega_{M}$, which gives $m_{d} \approx 0.15$ (Spergel et al. 2003), but this would imply 100\% efficiency in the conversion of halo baryons into disc baryons. It is known that a large fraction of baryons reside inbetween galaxies in clusters and a large fraction may remain in the halos of galaxies. For example Fukugita et al., (1998) give a value of $\Omega$ for stars in galactic disks of 0.0009 which leads to a value of $m_{d}=0.004$. They place most of the baryons in a hot gas around individual galaxies and groups. Silk (2004) says that only 50\% of the baryons predicted have been detected and of these 30\% reside in cold intergalactic gas, if the remainder were in galaxies then we would have $m_{d}=0.05$ which must be an upper limit. The mass-to-light ratios of some dwarf galaxies indicate values of $m_{d}$ even lower than this (Kleyna et al. 2004). MMW also conclude, after comparing their model with observations, that $m_{d}<0.05$ for typical disc galaxies. For the purpose of our simulation we will conservatively choose values of $m_{d}$ from a random uniform distribution between 0.01 and 0.05.

Running the simulation leads to the range of galaxy rotational velocities and scale sizes shown in fig. 2. At a given velocity width $\Delta V = 2 V_{c}$ the dark galaxies tend to have larger sizes. This can be understood by considering the scaling relations given in MMW (equations 12 and 13). Discs are larger and column densities lower if $m_{d}$ is small and/or $\lambda$ is large. The simulated data of fig. 2 can also be compared with observations by looking at fig. 4 of MMW. The candidate dark galaxy VIRGOHI21 has been included on Fig. 2b using the size (radius=7 kpc) and velocity width (200 kms) given in Minchin et al., (2006). 
Before we can compare the simulations with 21cm observations we have to consider what fraction of the gas in the disc of the galaxy is in the form of atomic hydrogen. We will assume that dark galaxies are metal poor and that they do not reach sufficiently high densities to form significant amounts of molecular hydrogen. To account for the contribution of helium we multiply the gas mass by a factor of 0.76 to obtain the HI mass. There is also the possiblity that the hydrogen in dark galaxies is ionised by the extragalactic UV background. Theoretical models  predict ionisation of atomic hydrogen layers that have column densities of a few $ \times 10^{19}$ cm$^{-2}$ (Maloney 1993). These theoretical predictions are supported by observations of the truncation of HI discs (van Gorkom et al. 1993) and star formation in tidal tails (Neff et al. 2005). Thus there may be a narrow range of column densities approximately between  $5 \times 10^{19}$ and $5 \times 10^{20}$ cm$^{-2}$ over which dark galaxies can exist and be detectable by 21cm surveys - they are not able to form stars, but are sufficiently optically thick to ionising radiation to retain a significant atomic component. In Fig. 3 we show the column density distribution of the stable dark galaxies produced by the simulation. Only a very small fraction (2\%) have column densities less than $5 \times 10^{19} $cm$^{-2}$. VIRGOHI21 has a central face-on column density of $3 \times 10^{19} $cm$^{-2}$ roughly inline with the above predictions.

The simulation predicts that about 80\% of stable discs have sufficiently high HI column densities that they can form stars. 
If all discs were still gaseous then about 2\% of the gas would be in dark galaxies. Given that those galaxies able to form stars will have done so we can adjust for the fraction of gas now turned into stars by dividing by $(1+\frac{1}{M_{HI}/L_{B}})$. Assuming that $L_{B}$ relates directly to stellar mass and taking the Milky Way value of  $M_{HI}/L_{B} \approx 0.1$ the model predicts that $\approx 15\%$ of the current HI in the Universe resides in dark galaxies. 

\begin{figure}
\psfig{figure=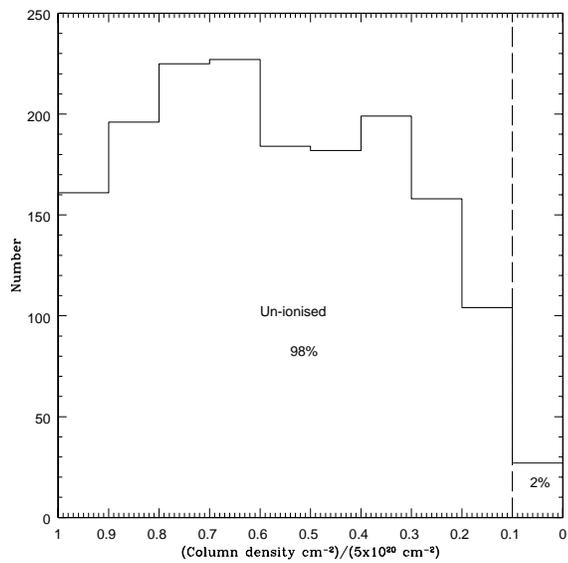,width=8cm}
\caption{The distribution of column densities (normalised by $\Sigma_{Crit}$) for stable dark galaxies. Only 2\% have low enough column densities to make them vunerable to ionisation.}
\end{figure}

We are now in a position to answer two of the three questions posed at the end of section 1. Firstly, the model predicts that dark galaxies $can$ exist with their relative numbers constrained by the range of $m_{d}$ and $\lambda$ that the Universe will actually allow at a given mass. In the model described above only a small fraction of the galaxies are dark. Can a dark galaxy with the properties of VIRGOHI21 exist ? The simulation certainly predicts that objects with scale sizes of tens of kpc and velocities of a few hundreds of km s$^{-1}$ can remain dark (Fig. 2). Relatively large objects like VIRGOHI21 appear to be viable, but they will be rare. In fig.4 we show histograms of the velocity widths and HI masses of the dark galaxies in the simulation. The majority of these objects have HI masses less than $\approx 10^{7}$ $M_{\odot}$ and velocity widths less than 40 km s$^{-1}$. We will show below that this makes dark galaxies very difficult to detect using currently available survey data.

\section{Simulating the HIPASS data}
Given the large amount of data available for galaxies detected by the HIPASS we will initially concentrate on comparing our simulation to the HICAT data. It is clear (Meyer et al. 2004, Minchin 2001) that the detection of sources within HI data cubes depends primarily on the peak flux of the source ($S_{p}$) and its observed velocity width ($W$).  This detection threshold appears to be $\approx 2.5\sigma_{rms}$ for the HICAT data (Meyer et al. 2004, Fig.9). Small values of ($W$) can only be detected if the the velocity channel width ($\delta V$) is also small and it is reasonable to assume that $2.5 \times \delta V$ is about the narrowest width that would be satisfactorly distinguishable from noise. Again this appears to be true for the HICAT data (Meyer et al. 2004, Fig.9). Large values of $W$ ($>1000$ km s$^{-1}$) may also be confused with baseline ripple. 

To obtain a peak ($S_{p}$) and an integrated flux ($S_{Int} \approx S_{p} W$) for our simulated galaxies we require a distance. This is done by choosing random numbers uniformly distributed over volume between distances of 0 and 176.4 Mpc (this corresponds closely to the velocity range of HICAT, 300 to 12,700 km s$^{-1}$ with $H_{o}=72$ km s$^{-1}$ Mpc$^{-1}$, Meyer et al. 2004). Each galaxy is assigned an inclination from a random number generator uniformly distributed over $\cos{i}$ where $i$ is the inclination of the galaxy to the plane of the sky. The observed velocity width is then taken to be $W = \Delta V\sin{i}$ with $W^{Min}=10$ km s$^{-1}$ to allow for the random component. Each galaxy that is able will convert some fraction of its HI mass into stars. HI selected samples of galaxies are observed to have values of $M_{HI}/L_{B}$ much higher than optically selected samples. For a subset of the HIPASS data that also has good quality optical photometry a mean value of $M_{HI}/L_{B}=1.5$ is found (Garcia 2005). For those galaxies that are able to form stars we reduce their HI content by $(1+\frac{1}{M_{HI}/L_{B}})$, which assumes $L_{B}$ relates directly to stellar mass.
For a comparison with the HICAT data we apply the selection criteria that $S_{p}> 2.5 \sigma_{rms}$ where $\sigma_{rms}=13$ mJy beam$^{-1}$ and $W>2.5 \delta V$ where $\delta V=13$ km s$^{-1}$ (See Fig. 9, Meyer et al. 2004). In Fig. 5 we show histograms of the velocities ($V_{Hubble}$), velocity widths $(W)$, peak fluxes ($S_{p}$) and integrated fluxes ($S_{Int}$) for our simulated galaxies. These correspond well with the HICAT data which are plotted as dashed lines on Fig. 5. The data are normalised to the same total number of galaxies in each sample. Qualatively the histograms appear to be remarkably good agreement when you consider the different origins of the two data sets. This is further substantiated in Fig. 6 where we show the bivariate distributions of the parameters of Fig. 5. Again Fig.6 should be compared with Fig. 9 of Meyer et al. 2004. From Fig. 5 and 6 we conclude that our model along with the HICAT selection criteria produces a reasonable realisation of the HICAT data.

\begin{figure}
\Large
{\bf a)}
\normalsize
\subfigure{\psfig{figure=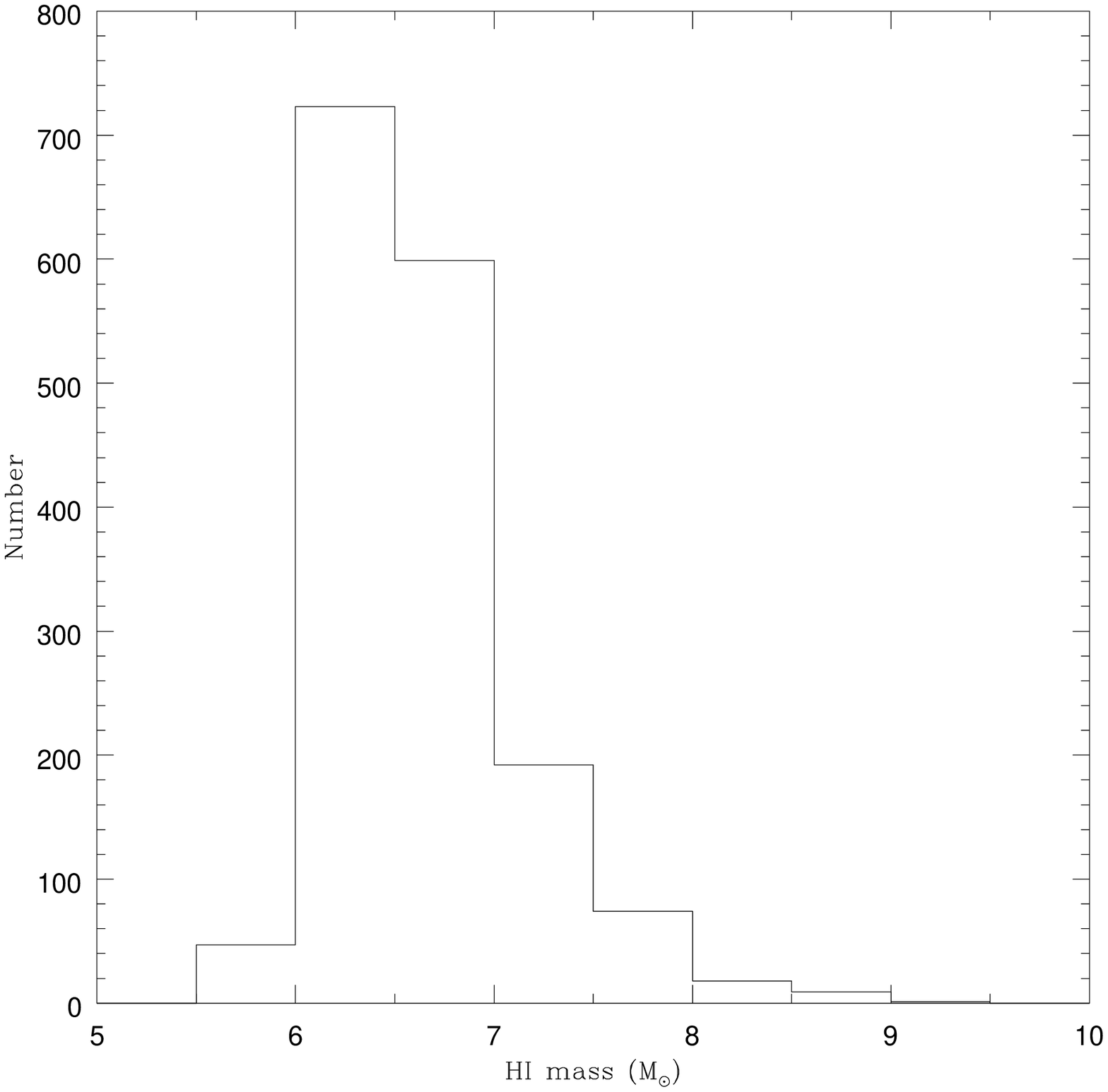,width=8cm}}
\Large
{\\ \bf b)}
\normalsize
\subfigure{\psfig{figure=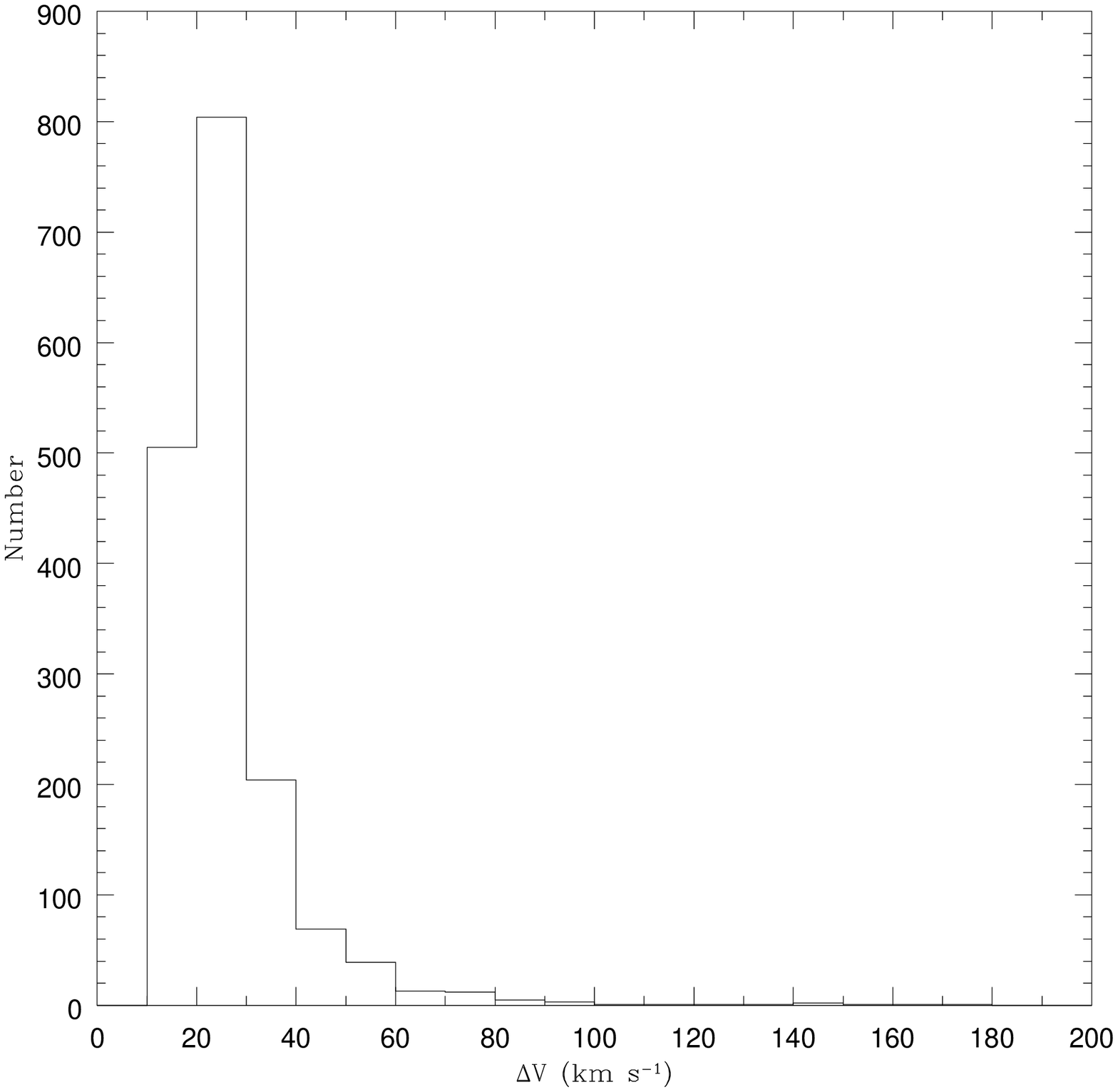,width=8cm}}
\caption{The simulation prediction for the distribution of a) HI masses and b) velocity widths for dark galaxies.}
\end{figure}

\begin{figure}
\subfigure{\psfig{figure=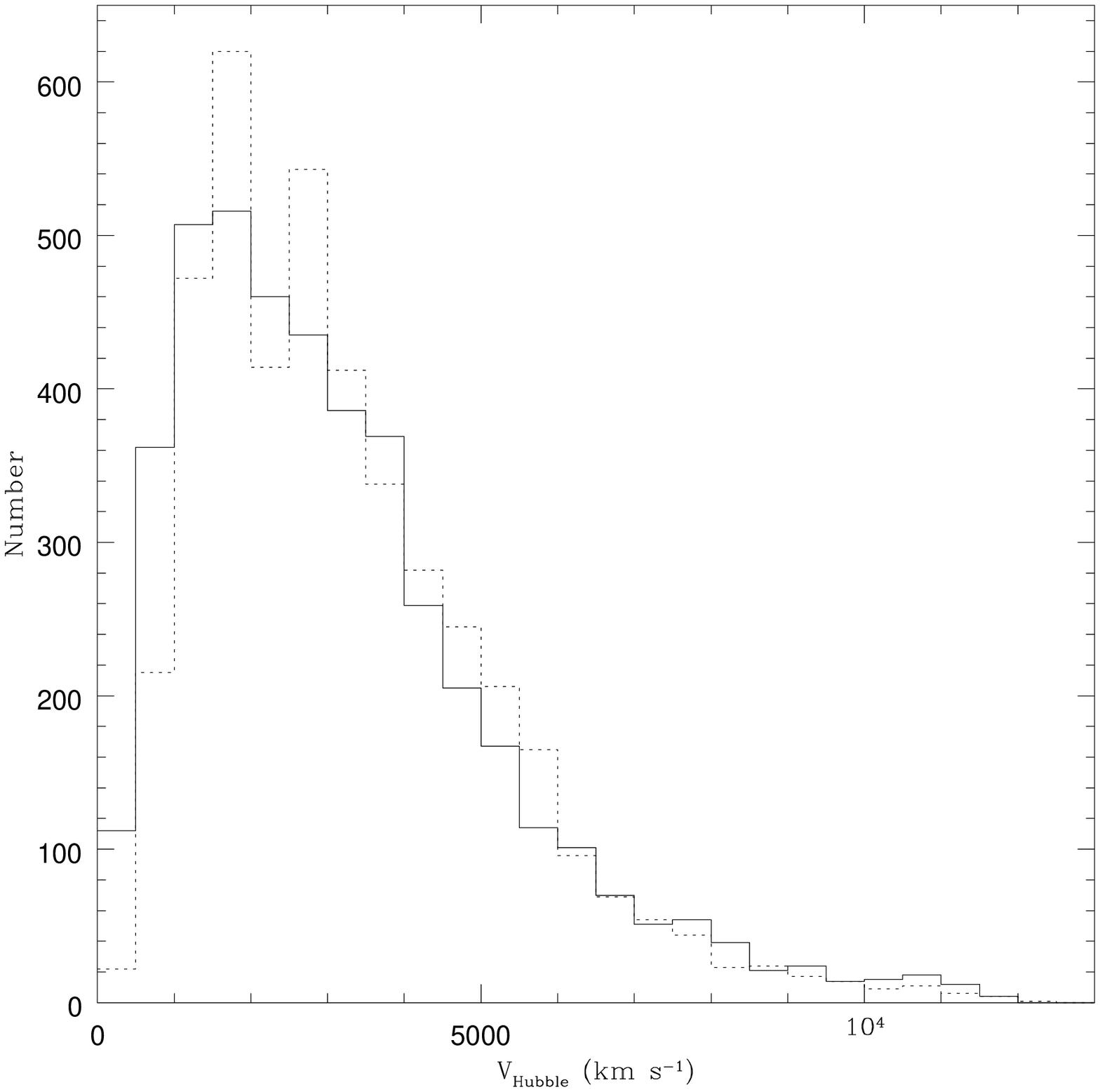,width=6cm}}
\subfigure{\psfig{figure=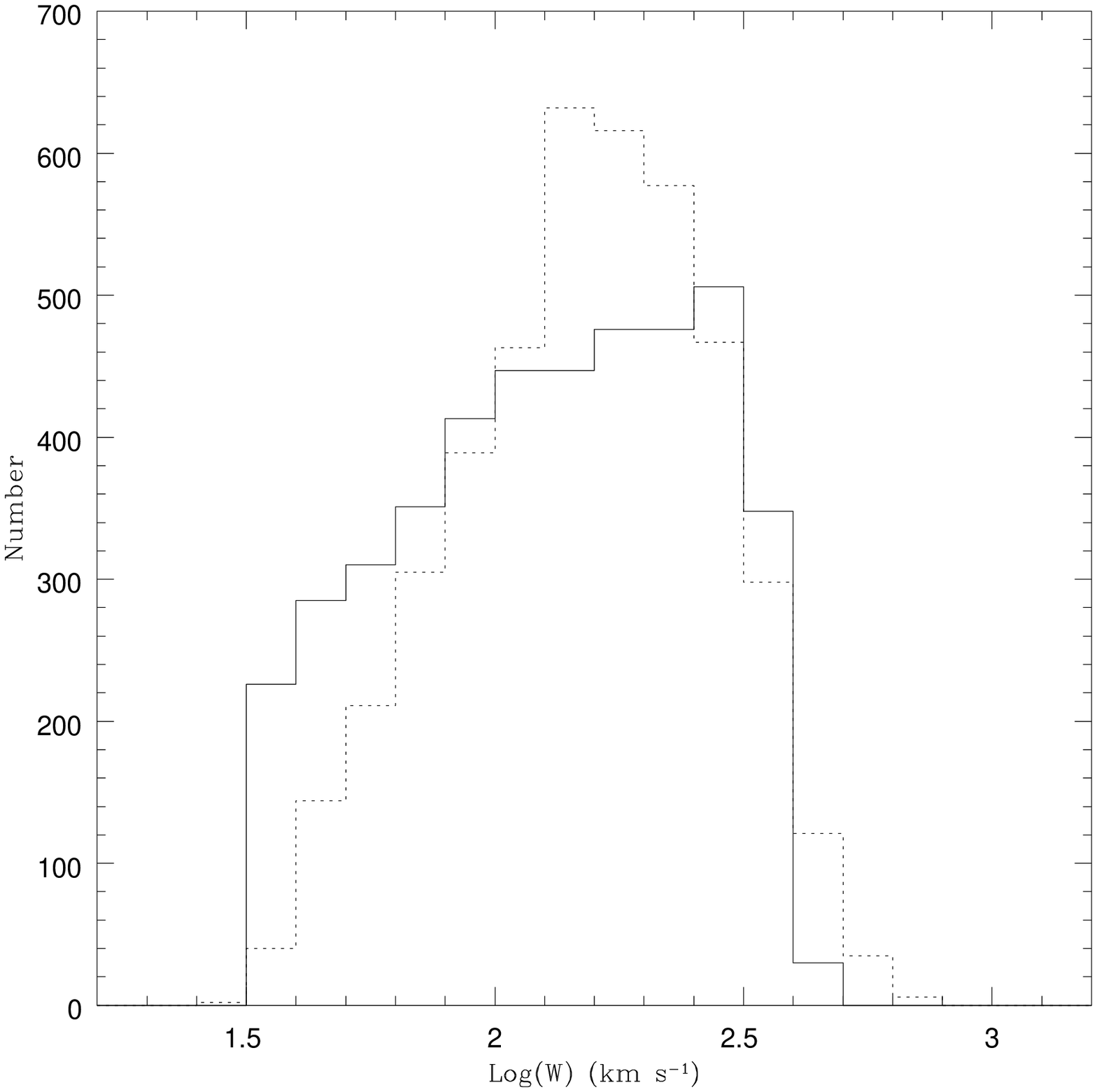,width=6cm}}
\subfigure{\psfig{figure=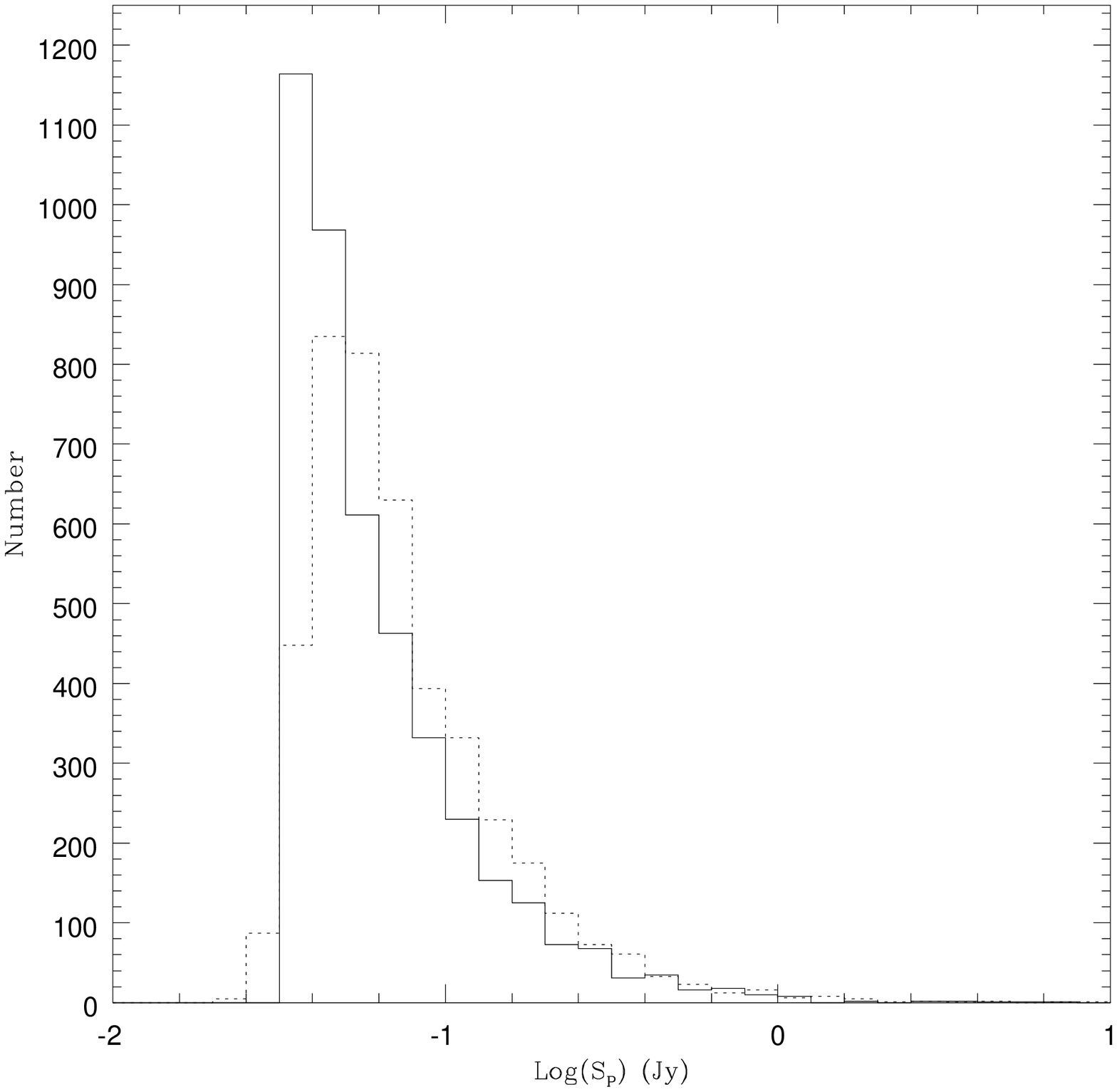,width=6cm}}
\subfigure{\psfig{figure=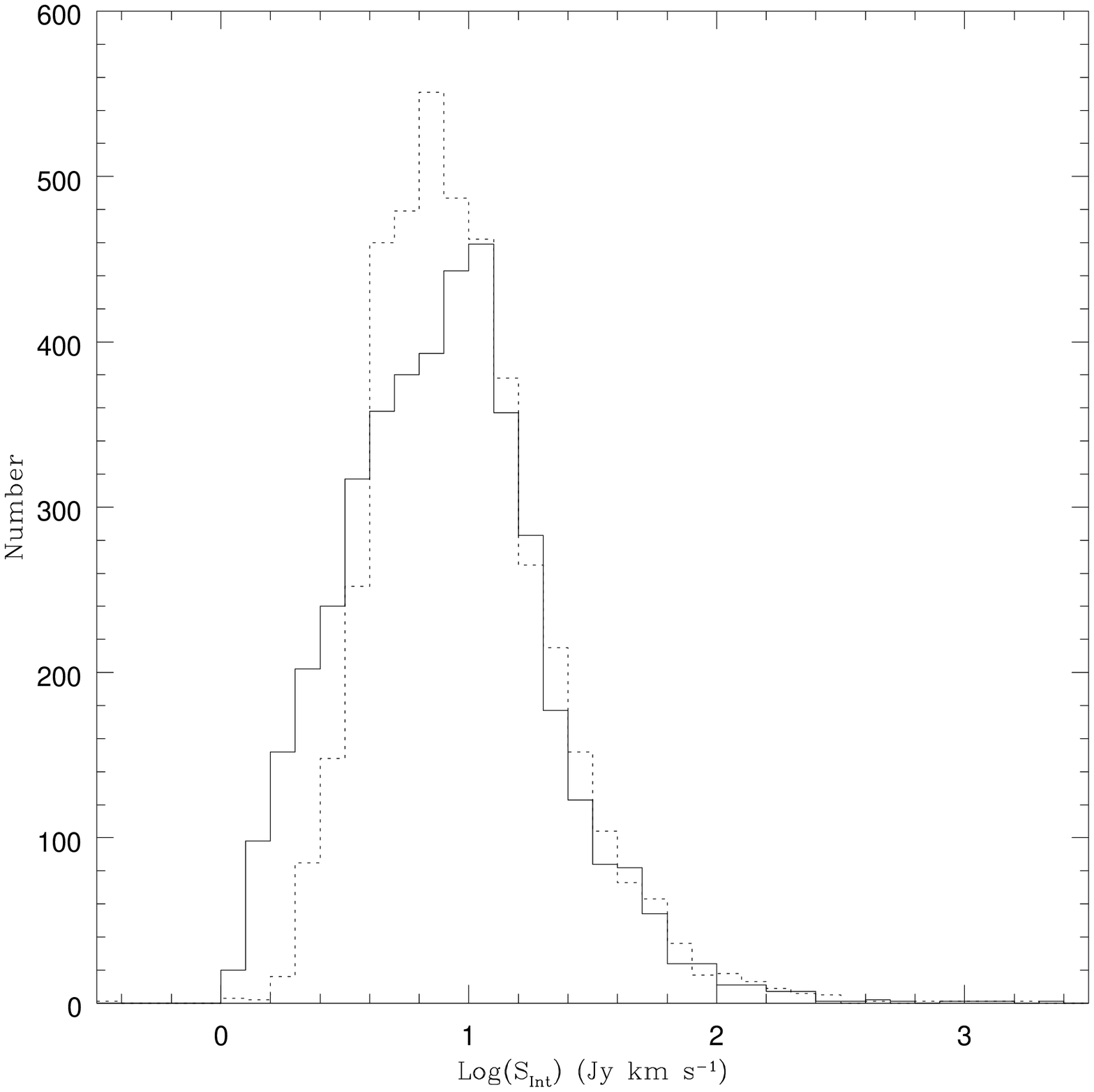,width=6cm}}
\caption{Histograms of the velocities ($V_{Hubble}$), velocity widths $(W)$, peak fluxes ($S_{p}$) and integrated fluxes ($S_{Int}$) for our simulated galaxies (solid line) and the HICAT data (dashed line). The histograms have been normalised to the same total number of galaxies.}
\end{figure}

\begin{figure}
\subfigure{\psfig{figure=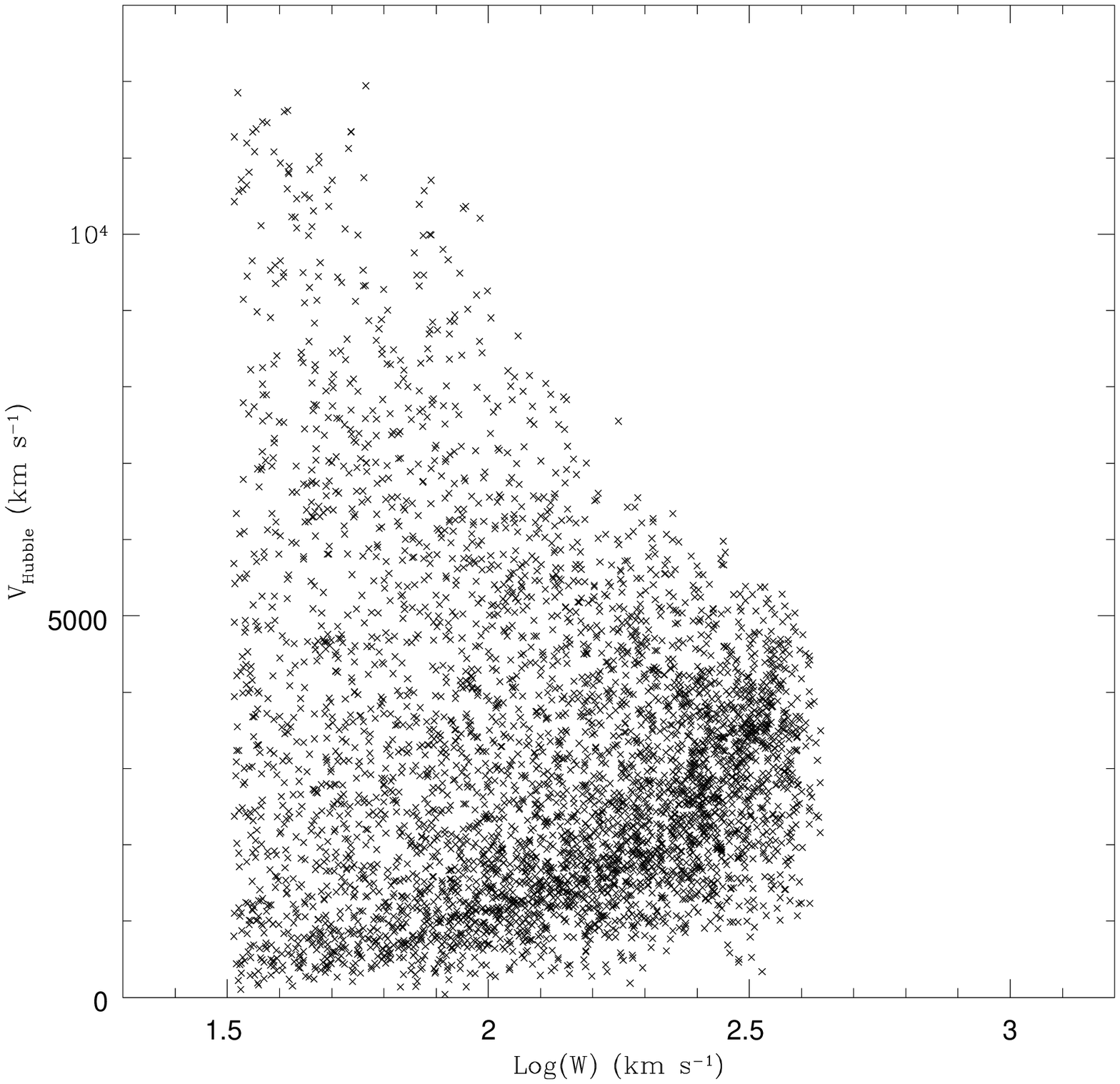,width=6cm}}
\subfigure{\psfig{figure=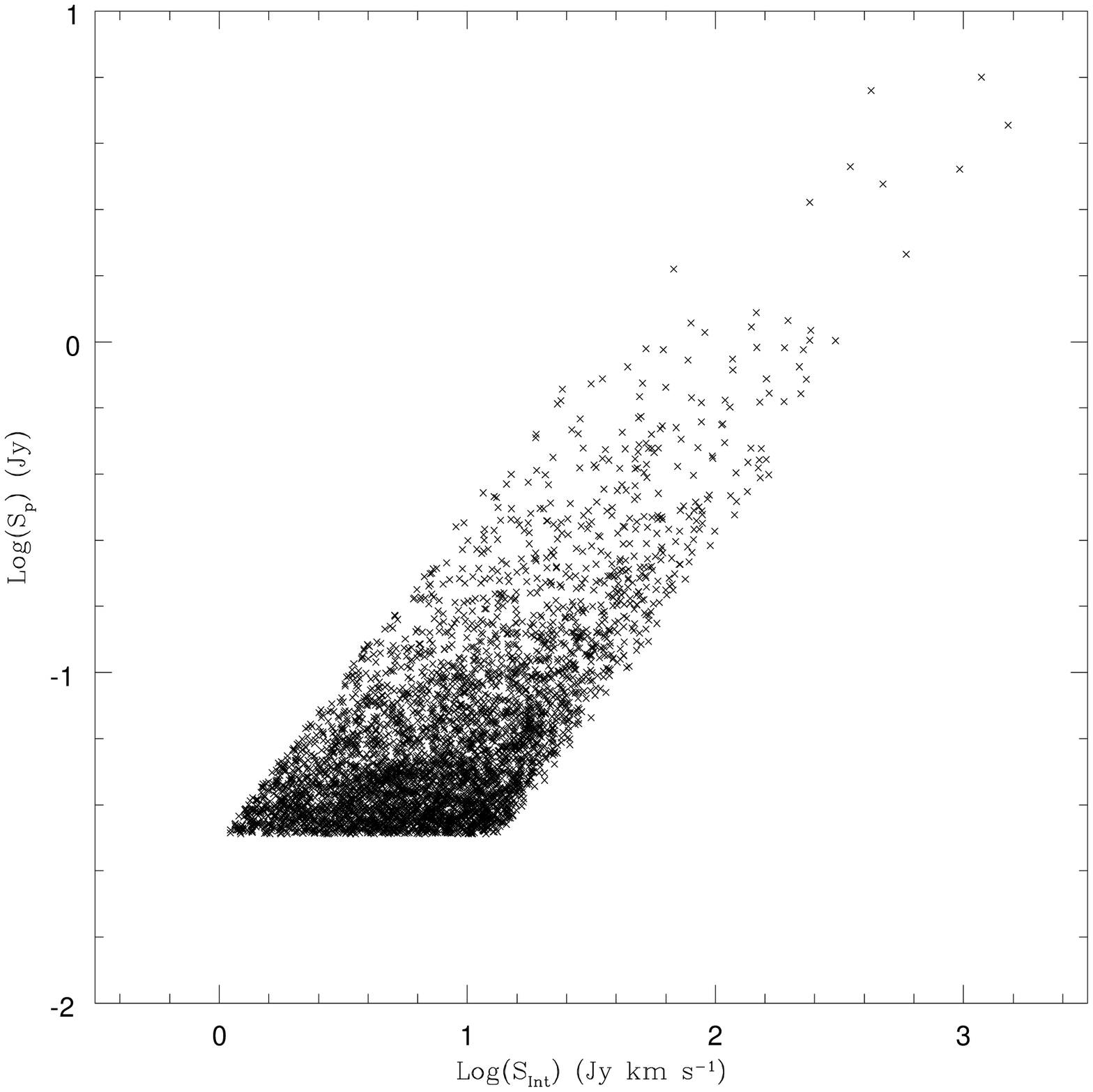,width=6cm}}
\subfigure{\psfig{figure=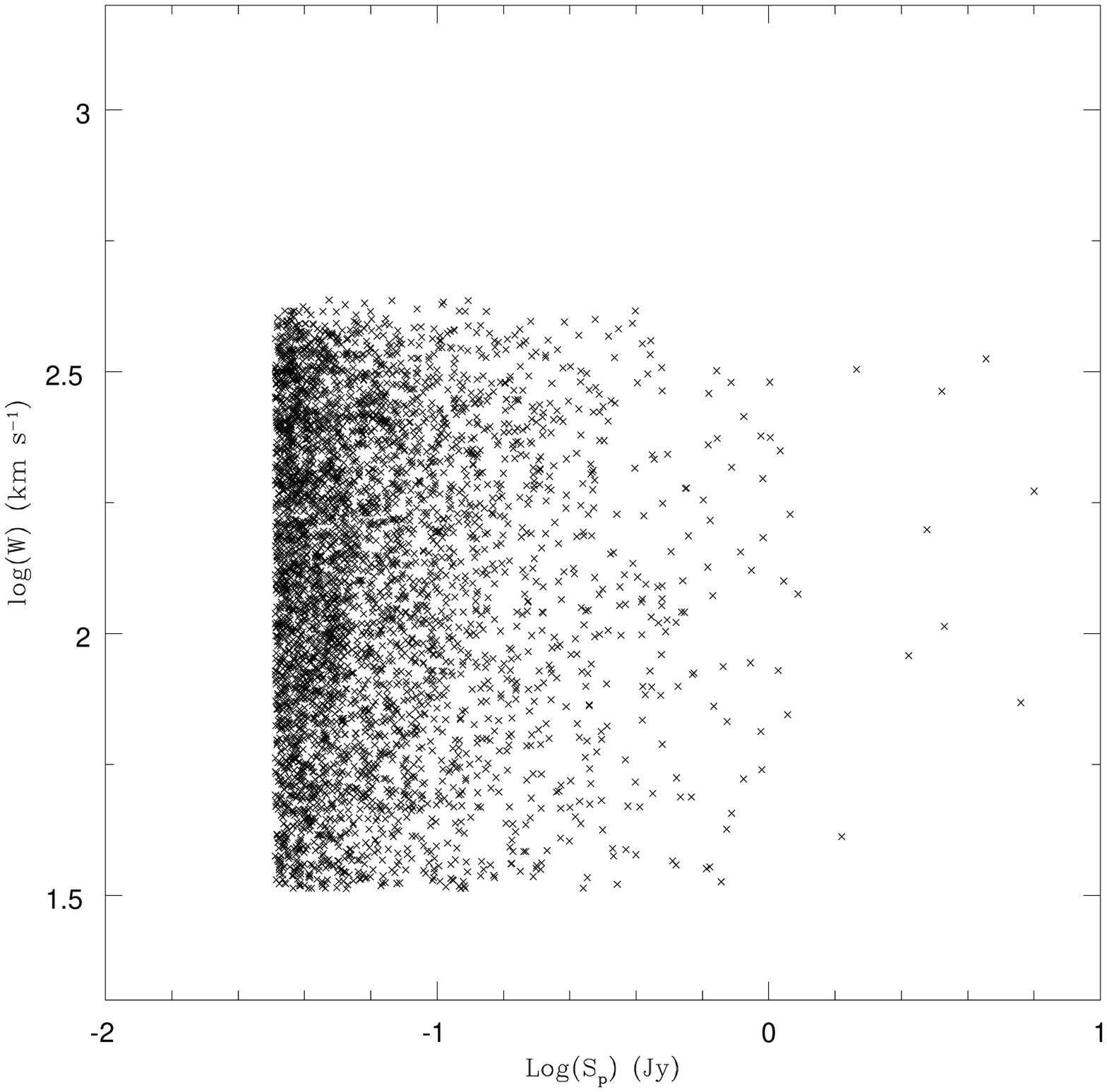,width=6cm}}
\subfigure{\psfig{figure=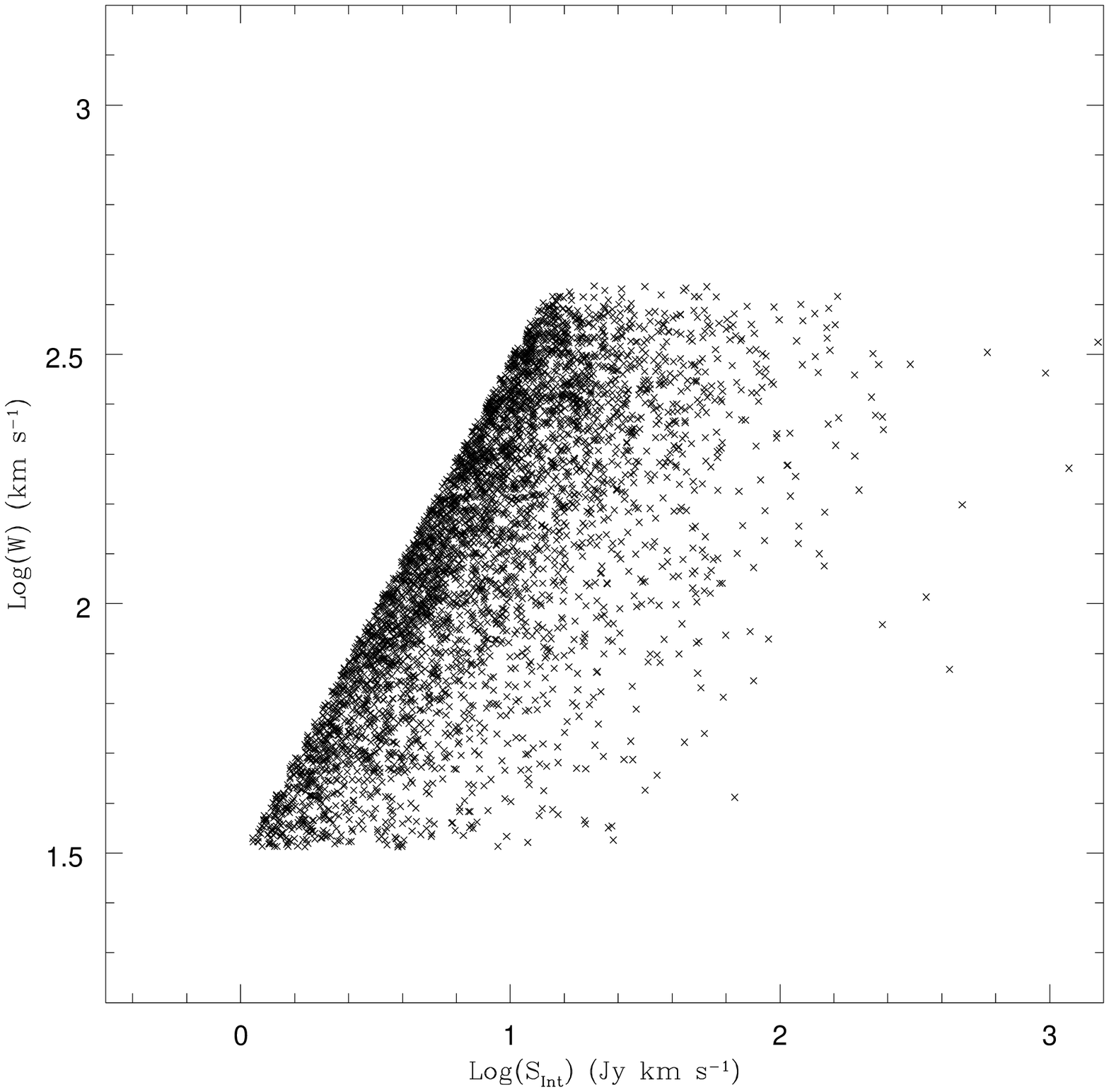,width=6cm}}
\subfigure{\psfig{figure=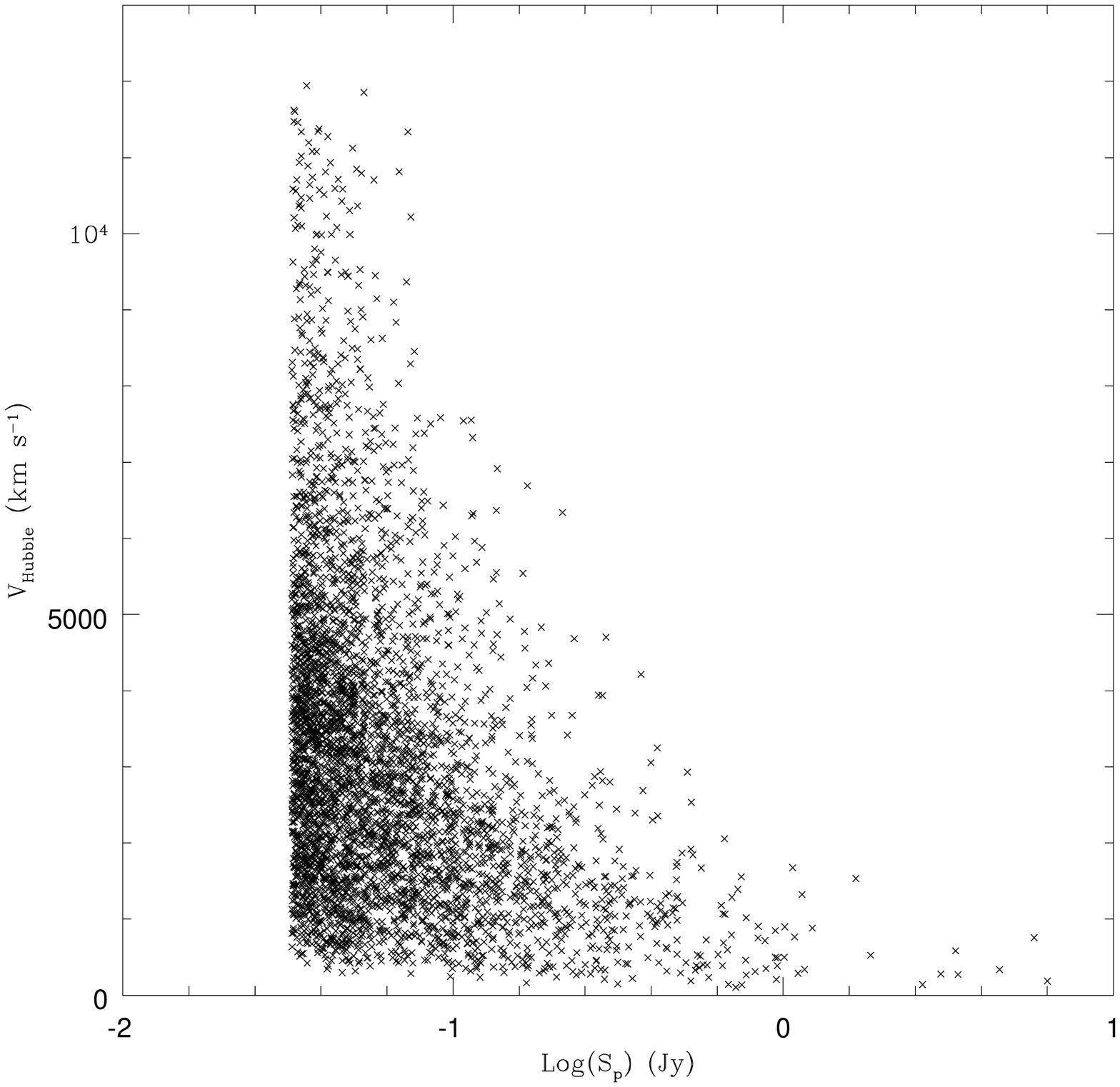,width=6cm}}
\subfigure{\psfig{figure=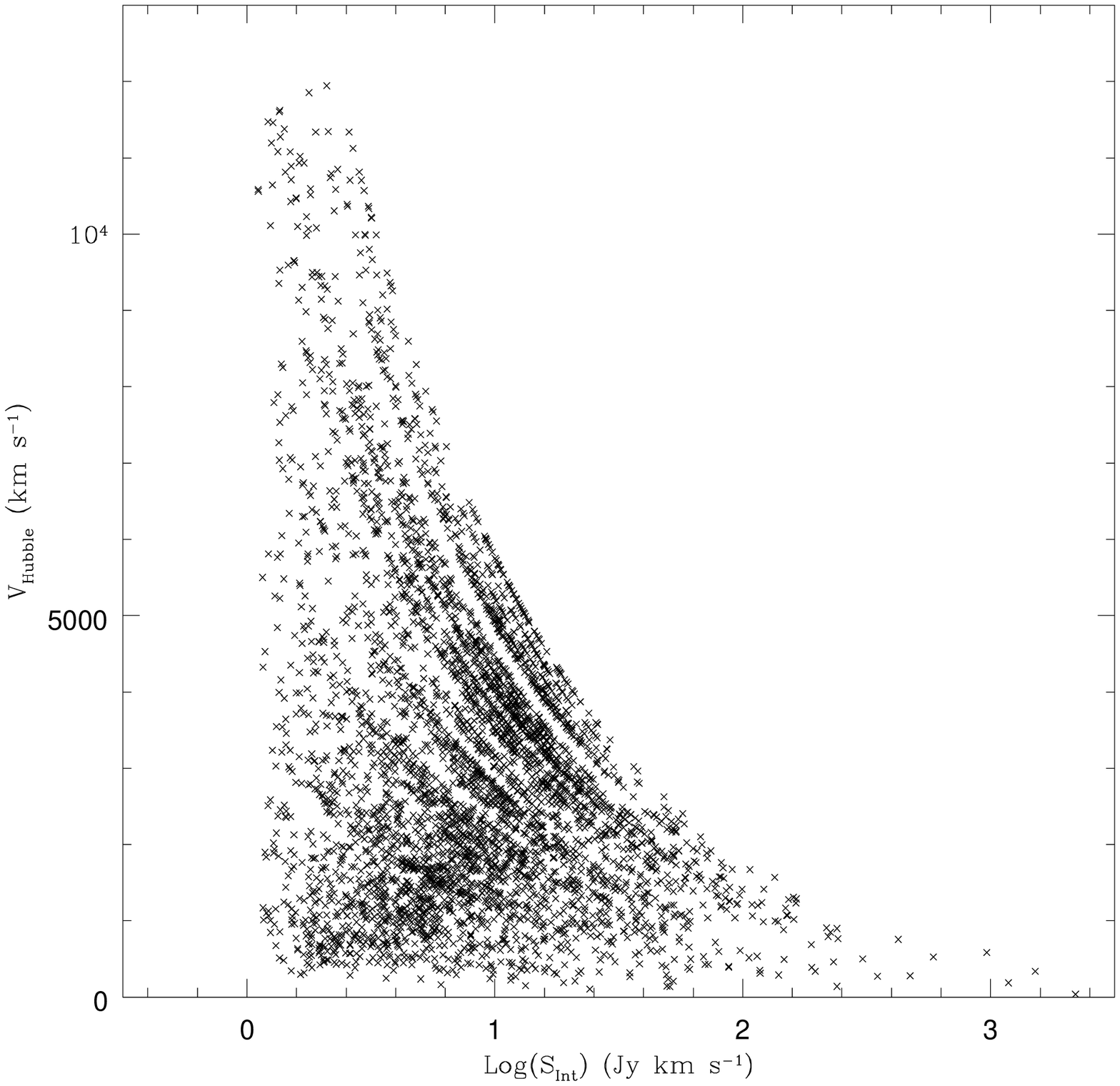,width=6cm}}
\caption{The bi-variate distributions of the various  parameters listed in fig 4 . This figure can again be directly compared to the HICAT data shown in fig. 9 of Meyer et al. 2004. }
\end{figure}

\begin{figure}
\subfigure{\psfig{figure=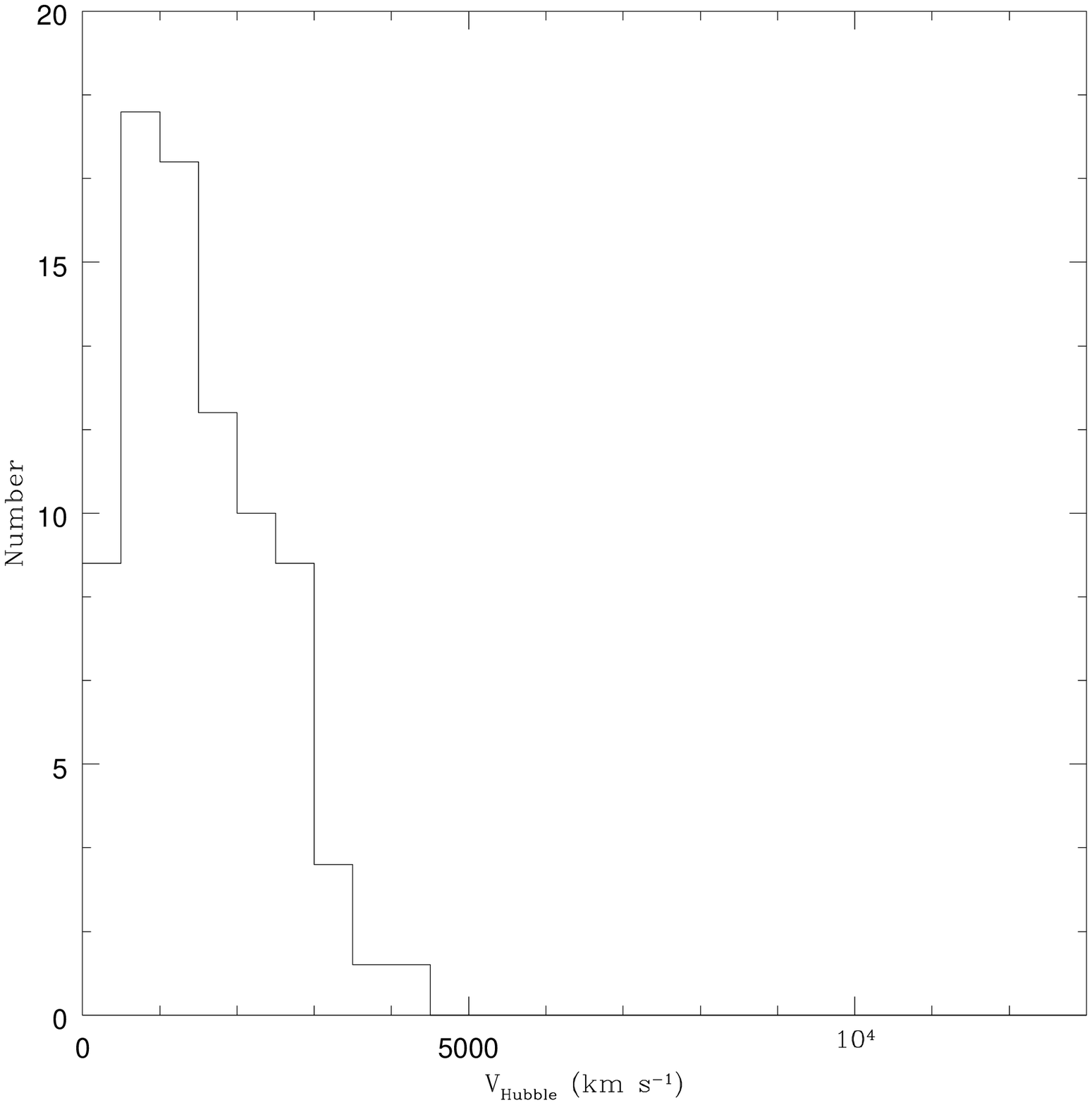,width=6cm}}
\subfigure{\psfig{figure=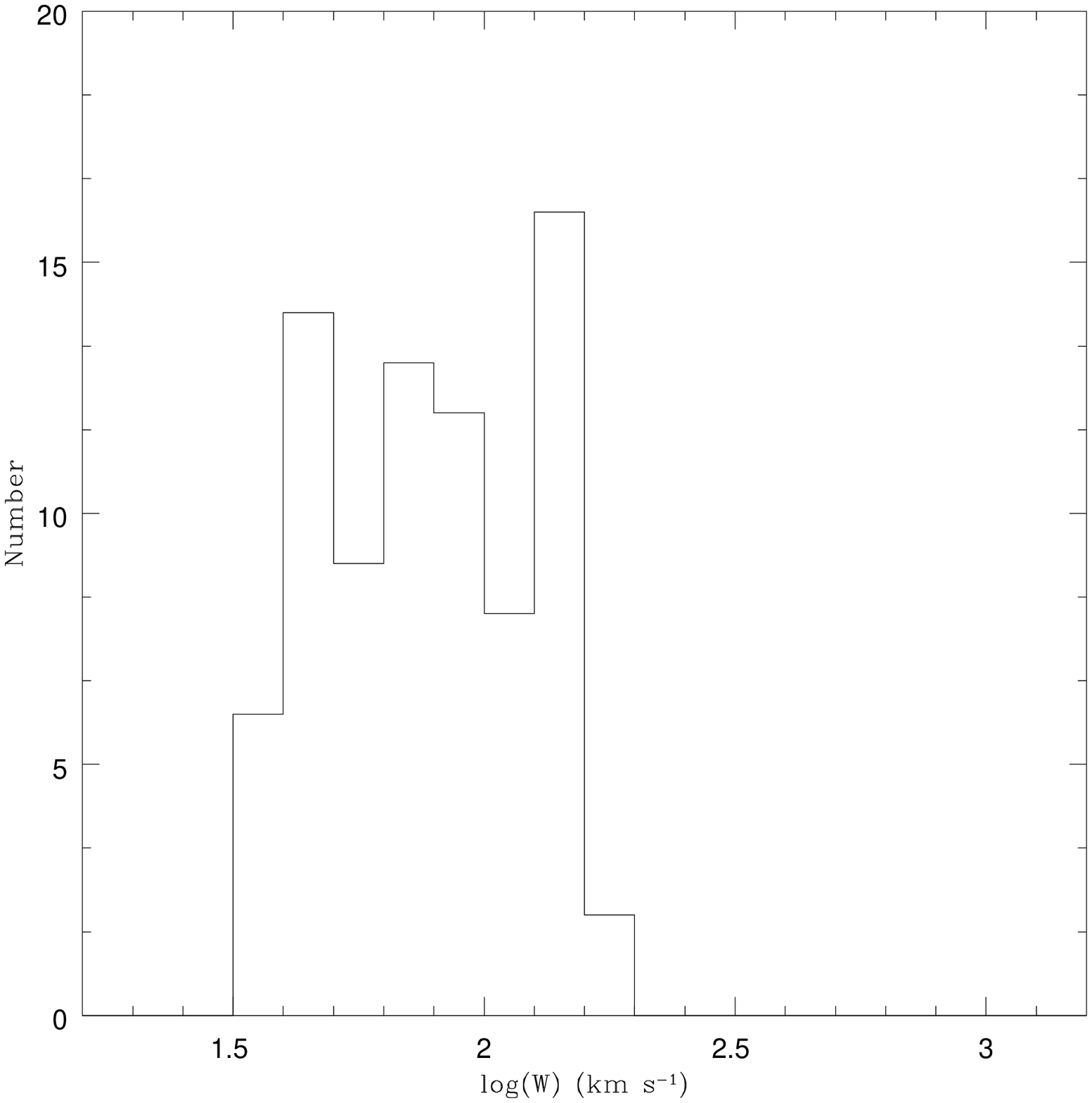,width=6cm}}
\subfigure{\psfig{figure=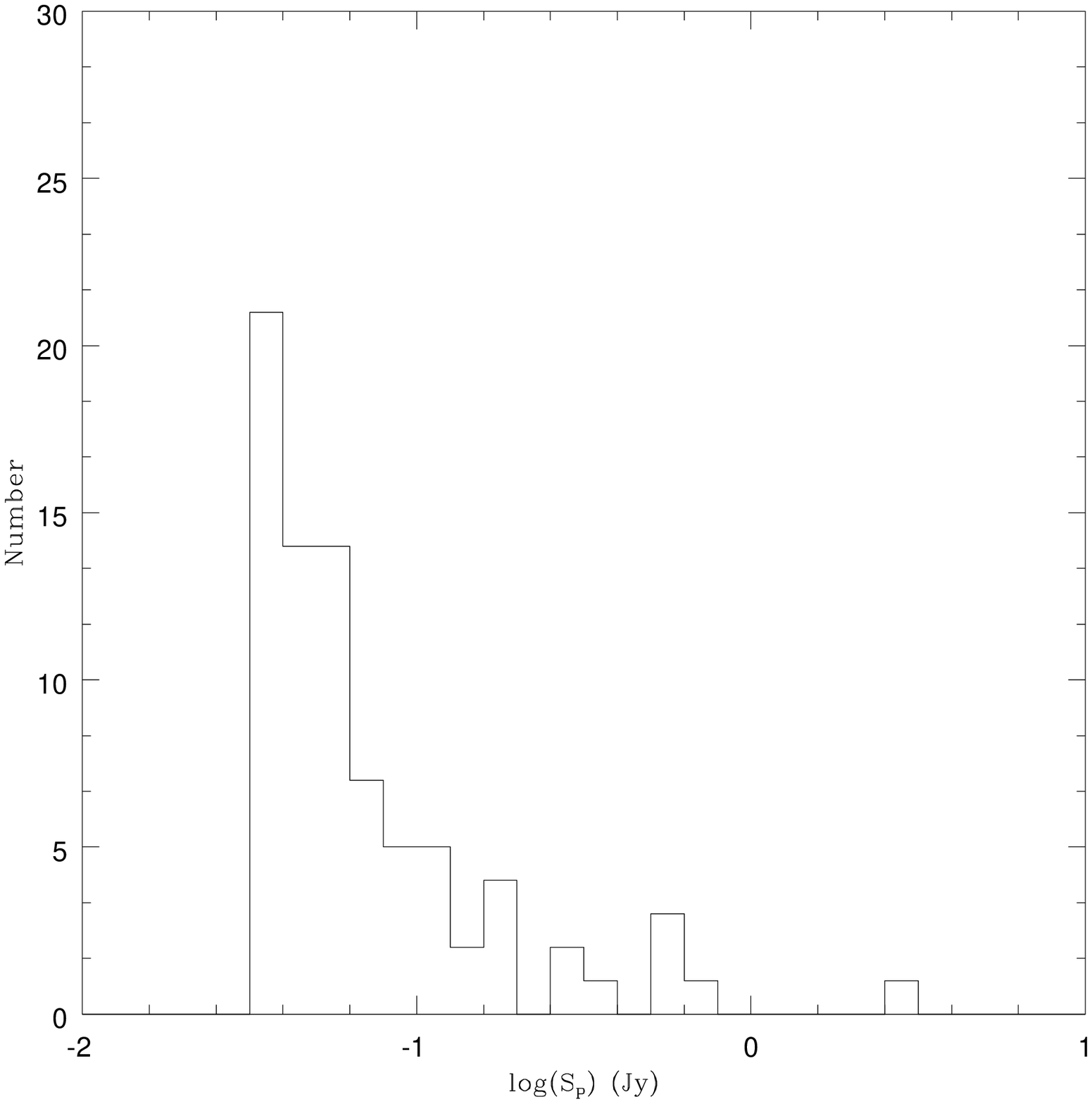,width=6cm}}
\subfigure{\psfig{figure=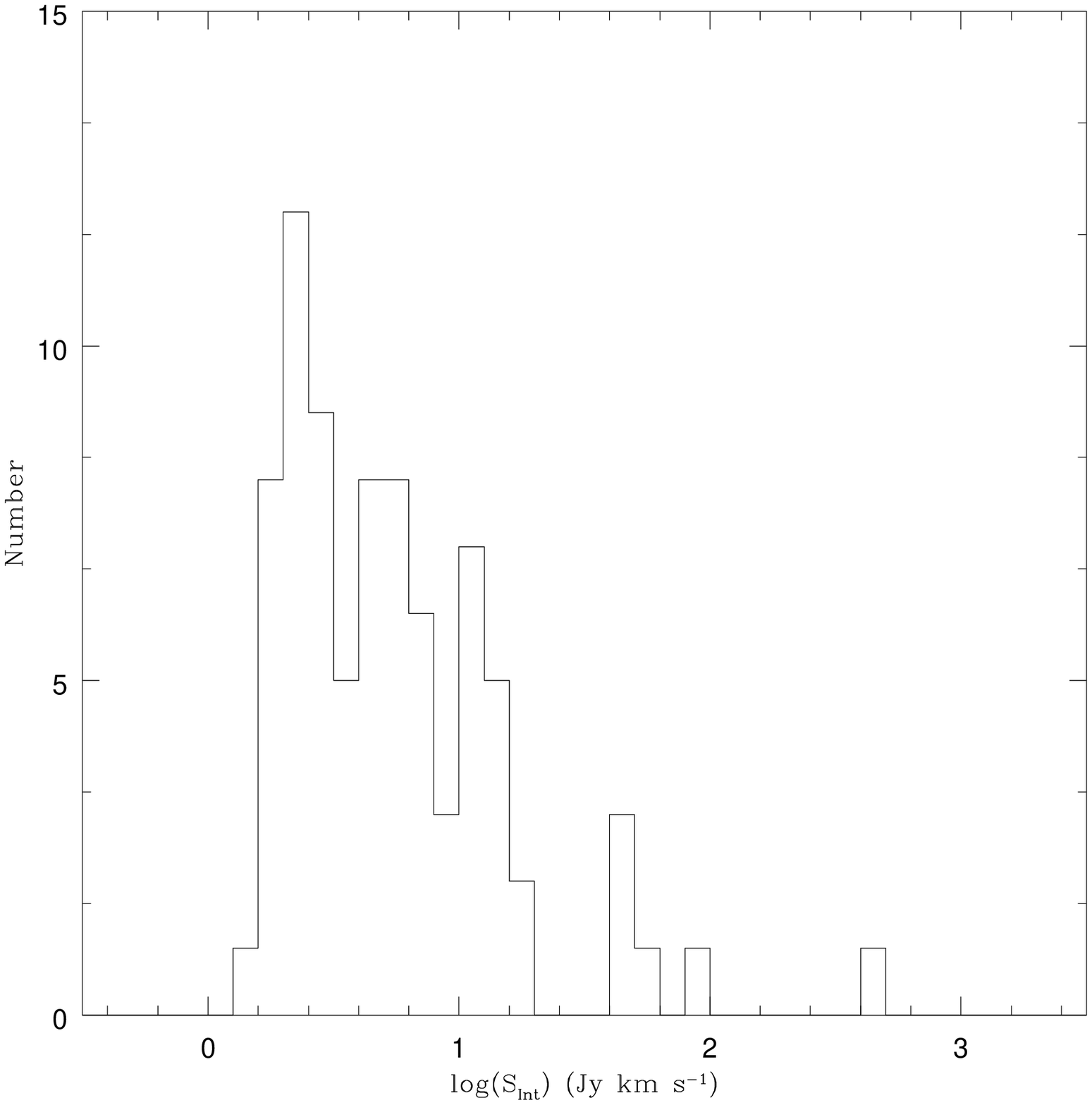,width=6cm}}
\caption{Histograms of the velocities ($V_{Hubble}$), velocity widths $(W)$, peak fluxes ($S_{p}$) and integrated fluxes ($S_{Int}$) for our simulated dark galaxies selected using the HICAT criteria. This figure can be directly compared to Fig. 5 and the HICAT data shown in Fig. 9 of Meyer et al. 2004. }
\end{figure}

The simulation predicts that just 0.3\% of the total HI detected in HICAT is in the form of dark galaxies. This compares with 15\% within the total simulated galaxy population i.e. we predict that many dark galaxies will go undetected by HICAT. Normalising the total numbers of galaxies in the simulation to the number in HICAT we predict that there should be $\approx 80$ dark galaxies in HICAT. In Fig. 7 we show the same histograms shown in Fig. 5, but this time for the dark galaxies only. There are two conclusions one can draw from Fig. 7. Firstly, the dark galaxies all reside within a relatively nearby 5000 km s$^{-1}$. Secondly, within the HICAT data the dark galaxies have similar distributions of properties as the galaxies that have formed stars.

Recently Doyle et al. (2005) carried out a search for optical companions of the 4315 HI sources in the HICAT. Their conclusion is that there are no 'isolated' dark galaxies.  The statistics of what was found is as follows: there are optical counterparts for 3618 HI sources (84\%), this includes what they describe as 'good guesses', 972 of these (23\%) are described as 'multiple possible matches', 2512 (58\%) have confirmed optical velocities, 216 (5\%) have no obvious optical galaxies present (most of these are in crowded fields, along the Galactic plane or have high extinction). They also say that some of their optical matches disagree with the optical matches given in the HIPASS bright galaxy catalogue (Koribalski et al. 2004). Doyle et al. define a dark galaxy '..as any HI source that contains gas (and dark matter) but no detectable stars, and is {\it sufficiently} far away from other galaxies, groups or clusters such that a tidal origin can be excluded'. It is also important to note that the Parkes beam is about 14 arc min across and that Doyle et al. assume a 'velocity match' if HI and optical velocities agree to 400 km s$^{-1}$. 

In practice there is no reason to expect that dark galaxies will be any less clustered than luminous galaxies. In that case it might be all too easy to misidentify a dark galaxy with a bright candidate galaxy in the same cluster or group, or even, lacking optical velocities, with bright galaxies in the background or foreground. A rough calculation suffices to show that this is the case.
Take the most optimistic situation, i.e. for galaxies nearby ($V_{Hubble}<1000$ km s$^{-1}$ or distance modulus of 30.7) where it is hardest to make such a misidentification. At 1000 km s$^{-1}$ HIPASS can find sources with as little as $10^{8}$ $M_{\odot}$ of HI. Provided one is prepared to allow identification with objects which have $M_{HI}/L_{B}$ as high as 10 - as Doyle et al. in fact do, then identifications can be made with galaxies with luminosities no greater than $10^{7}$ $L_{\odot}$ i.e. with dwarfs as faint as $M_{B}=-12.5$ or background galaxies as faint as $m_{B}=18$. Shanks et al. (1991) find $10^{5.1}$ galaxies str$^{-1}$ mag$^{-1}$ at $m_{B}=18$ which means there will on average be more than one such object within 7 arc min of any radio position chosen at random on the sky. The situation is of course much worse for more distant galaxies. For galaxies at distances greater than about 1000 km s$^{-1}$ the situation is not clear-cut even if an optical redshift is available. If dark galaxies are clustered with bright galaxies and identifications are permitted with optical sources up to 7 arc min away and within 400 km s$^{-1}$, as in Doyle et al., then an optical redshift will not uniquely identify the HI source - there will be just too many nearby companions (Disney et al. 2005).  

We do not believe that the Doyle et al. (2005) identification statistics and selection criteria rule out our predicted 80 dark galaxies within HICAT.

\section{Simulating the other surveys}
We can now use the parameters of the other blind HI surveys listed in table 1 to predict the numbers of dark galaxies they are likely to detect. We use the same very conservative selection criteria i.e. $S_{p} > 2.5 \sigma_{rms}$ and  $W > 2.5 \delta V$  and carry out the simulation over the volume defined by $V^{Max}$. In table 2 we give the results of these simulations. It is clear that the predicted percentages of dark galaxies within all previous surveys is very low. But are the predicted numbers consistent with previous HI surveys?

\begin{table}
\begin{center}
\begin{tabular}{l|c|c|c}
Survey  &  \% Predicted   & Total HI    & Predicted Number of\\
        &  Dark Galaxies  & Detected & Dark galaxies in Survey  \\ \hline
Henning &  3              & 37       & 1   \\
AHISS   &  5              & 66       & 3  \\
ASS     &  3              & 75       & 2  \\
ADBS    &  2              & 265      & 5  \\
HIJASS  &  2              & -        & -   \\
HIPASS  &  2              & 4315     & 80 \\
HIDEEP  &  6              & 129       & 8    \\
ALFALFA &  4              & 20000    & 800  \\
VIRGOHI &  6              & 31       & 2   \\
AGES    &  23             & -        & -
\end{tabular}
\end{center}
\caption{The predicted percentage of dark galaxies in various blind HI surveys. HIJASS does not have complete identifications of sources, the ALFALFA survey is still being carried out and the AGES survey is yet to start. The 20,000 detections for ALFALFA is a prediction from initial observations. The numbers for the VIRGOHI and AGES surveys are for a galaxy population simulated at about the distance of the Virgo cluster (16 Mpc).}
\end{table}

The Henning (1995) survey is predicted to contain just one dark galaxy out of the 37 HI detections. Given that the survey also primarily covered the ZOA the results place no real constraints on the existence of dark galaxies. 

The AHISS detected 66 galaxies but some of the area covered was again in the ZOA. There is also a variation of sensitivity of about a factor of ten across the beam. Zwaan et al. (1997) state that they have a detection threshold of $5\sigma$ which is twice that used to produce the numbers in table 2. Schneider et al. (1998) say that this is actually $7\sigma$ if you consider the way the data is reduced. If one includes only those detections that are within the telescope main beam and outside the ZOA then there are just 26 detections and the prediction of table 2 becomes just one dark galaxy. 

Of the 75 detections in the ASS 35 are listed in major catalogues and all but one was detected in CCD follow up. The one remaining galaxy must have a surface brightness fainter than 25 $B\mu$ and fits well with our prediction of two dark galaxies.

Part of the ADBS survey was again in the ZOA. Of the 81 new detections 11 had no optical counterparts after inspection of the digital sky survey. This certainly leaves open our prediction of 5 dark galaxies in this survey.

HIDEEP (Minchin et al. 2003) made 173 HI detections of
which 129 were in the region for which optical follow-up was obtained
(Minchin et al. 2004).  Of these, only 107 could be
uniquely identified with galaxies and only 87 were `secure' in that they
had either matching optical redshifts or HI interferometry.  Thus almost a
third of the HI detections within the optical follow-up area do not have
secure optical associations.

Given the above selection criteria the maximum distance an object of mass $M_{HI}$ can lie at and still be in the sample is proportional to $(\sigma_{rms} \delta V)^{-1/2}$. The volume corresponding to this distance is
\begin{equation}
Vol \propto A (\sigma_{rms} \delta V)^{-3/2}
\end{equation}
where $A$ is the angular area covered by the survey. It is typically found that $\sigma_{rms}$ decreases as $t_{exp}^{1/2}$ where $t_{exp}$ is the integration time per pointing (Minchin 2001). In this case the volume surveyed is proportional to $A t_{exp}^{3/4}$. This latter relation shows that to find low mass sources in a survey selected in this way it is more efficient to increase the area of the survey than the integration time per point. This explains why HIPASS is predicted to have so many dark galaxies, even though its sensitivity is low and its velocity resolution is poor - it has a large area coverage. The currently in progress ALFALFA survey wins on all counts as it has both large area coverage, sensitivity and velocity resolution. Even so the fraction of dark galaxies is still low making them difficult to find amongst the optical galaxies and requiring good optical follow up. Note that the stated (table 2) total of 20,000 galaxy detections for ALFALFA is a prediction taken from Giovanelli et al, (2005).

Finally, we note that the VIRGOHI and the AGES surveys (table 2) are different types of surveys to the others. In the main both of these surveys were specifically designed to identify HI objects at one particular distance. In table 2 to obtain the prediction for the VIRGOHI and AGES surveys we have simulated the galaxy population at just one distance (16 Mpc). It is clear that the fraction of dark galaxies that may be detected is much higher if this observing strategy is adopted - particularly so for the AGES survey which is able to detect the low mass small velocity width objects that are the characteristics of many of the simulated dark galaxies (Fig. 4).  It may be no accident then that the candidate dark galaxy VIRGHI21 was detected in a survey like this and the predicted number of two dark galaxies is fully consistent with our detection of one. We have already predicted that these objects are rare and so given that dark and light galaxies cluster together our chances of finding them are maximised when we look in places where there are lots of optical galaxies.

\section{The effect of altering critical parameters}
In section 2 we described our model and said that for our initial best guess parameters we predict that about 20\% of all galaxies are dark and about 15\% of the total HI currently in galaxies will reside in dark galaxies (using $M_{HI}/L_{B}=0.1$). Of the number of adjustable parameters in the model the low mass slope of the mass function ($\alpha$) and the fraction of the mass that resides in the disc ($m_{d}$) have the largest influence on our conclusions. For example if $\alpha=-1.0$, all other things remaining unaltered, then only 9\% of galaxies are dark and about 1\% of the HI currently in galaxies is in dark galaxies. If $\alpha=-2.0$ then $\approx$25\% of galaxies are dark, and they will contain $\approx$50\% of the HI. More dramatically changing $m_{d}$ to 0.004 (the value taken from Fukugita et al. (1998) for stars in galactic discs, section 2) the model predicts that $\approx$74\% of galaxies are dark and that $\approx$67\% of galactic HI is in dark galaxies ($\alpha=-1.5$). Applying the HICAT selection criteria to the above simulated samples of galaxies we predict that for $\alpha=-1.0$ 1\% of the HICAT galaxies would be dark (43 galaxies). For the case $\alpha=-2.0$ 6\% would be dark (259 galaxies). Fixing $m_{d}=0.004$ ($\alpha=-1.5$) leads to a predicted 34\% of galaxies being dark in HICAT (1467 galaxies). From the Doyle et al. (2005) paper $m_{d}=0.004$ is probably ruled out. In our view $\alpha=-2.0$ is still possible, but less likely than a shallower faint end slope to the mass function. 

The value of $m_{d}$ is crucial to understanding the dark galaxy issue. We, from lack of any better knowledge have chosen, in our simulations, to make $m_{d}$ a random variable uniformly distributed over 0.01-0.05. Others have chosen differently and some come to different conclusions. Verde et al. (2002) make $m_{d}$ a function of halo mass with it taking a value of 0.026 at $M=10^{10}$ $M_{\odot}$ and 0.07 at $M>10^{12}$ $M_{\odot}$ (see their Fig. 2). This functional form of $m_{d}$ is well motivated by the idea that smaller haloes may lose a larger fraction of their baryons during the first stages of galaxy formation. Verde et al. use their model to explain the difference between the predicted CDM faint end slope of the dark matter mass function and the observed luminosity function of galaxies. In addition they use the Toomre criterion to assess whether stars will form, rather than our fixed column density threshold. They conclude that '...a large fraction of low-mass dark matter haloes may form Toomre-stable discs....Such halos would be stable to star formation and therefore remain dark.'. This is quite the opposite conclusion to which Taylor and Webster (2005) come. They create a very similar model to both ours and that of Verde et al. but study in detail how the gas may cool to form stars in the presence of an ionising background. Their conclusion is that almost all discs formed in their simulations have sufficiently high column densities to self-shield the hydrogen so that $H_{2}$ can form, cool and produce stars. The reason they come to this conclusion is that they take $0.05<m_{d}<0.1$, but as we previously showed in Fig. 1 such large values of $m_{d}$ lead to the inference that dark galaxies are extremly rare or do not exist. Given that $\Omega_{Baryon}/\Omega_{M}$ gives $m_{d} \approx 0.15$ (Spergel et al. 2003) a value $m_{d}=0.1$ would imply an implausibly high retention of baryons within halos (contrary to observations of inter-galactic hot gas). Although Taylor and Webster reference MMW for the values of $m_{d}$ that they use, MMW specifically say that $m_{d}<0.05$. If Taylor and Webster had used more realistic values of $m_{d}$ they would surely come to the same conclusion that Verde et al. and we do.

\begin{figure}
\psfig{figure=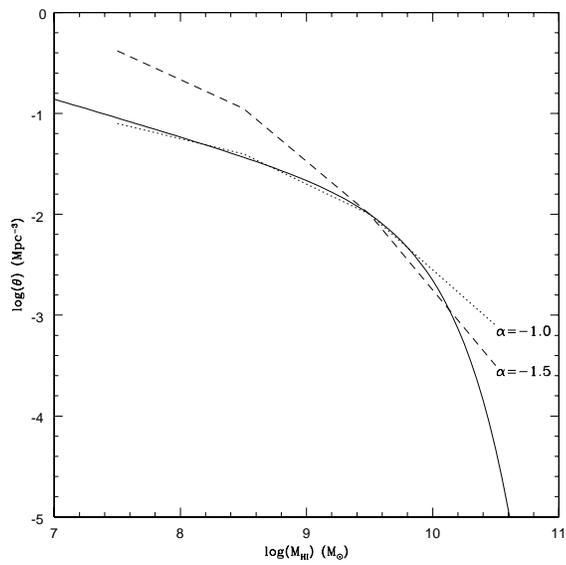,width=8cm}
\caption{The HI mass function of all those simulated galaxies selected using the HICAT criteria for two different faint end slopes (dashed and dotted lines) compared to the observed HI mass function derived from HICAT by Zwaan et al. (2005).}
\end{figure}

Finally, we can consider the HI mass function predicted by our model and compare this with that obtained from the HICAT data (Zwaan et al., 2005). We have selected galaxies from our simulations according to the HICAT criteria described above and then applied the volume corrections to obtain the HI mass function of the simulated data. The result is shown in Fig. 8 where we compare the results of two simulations ($\alpha=-1.0$ and $\alpha=-1.5$, see above) with the HI mass function obtained by Zwaan et al. (2005). All the curves have been normalised to a value of -2 at $\log(M_{HI})=9.5$). The first thing to note is that the measured low HI mass slope obtained from the simulations is steeper than the input mass function (-1.3 for $\alpha=-1.0$ and -1.6 for $\alpha=-1.5$). The input mass spectrum has been modified by the requirement that galaxies are stable (eq. 8), that they satisfy the selection criteria and because some, but not all, galaxies have their HI mass reduced by star formation - relatively more initially high mass galaxies have their HI depleted in this way. A problem with each of the simulations is that they predict more high mass galaxies than are observed. The simulation with $\alpha=-1.0$ provides a surprisingly good fit to the low mass end of the HICAT mass function. Given that the numbers of dark galaxies in HICAT are low and the good fit to the observed HI mass function for an intrinsic $\alpha$ of -1.0 Occam's razor suggests, contrary to the favoured galaxy formation models, that the dark matter mass function is flat. 

\section{Summary}
In this paper we have described a model of the galaxy population of the Universe. We have used this model to firstly infer that dark galaxies are a possible outcome of the galaxy formation process. Secondly, we have shown that the simulations predict the existence of dark galaxies including galaxies with similar properties to the candidate dark galaxy VIRGOHI21. Thirdly, we have described how previous HI surveys have placed few if any constraints on the types of dark galaxies produced by our simulation. None of these surveys have been able to rule out the existence of dark galaxies because: 
\begin{enumerate}
\item Dark galaxies are rare compared to galaxies with stars.
\item Dark galaxies can have the same ranges of size and mass as do optical galaxies, but a larger fraction of them are small and of low mass. This makes them difficult to find.
\item Previous surveys have not had sufficient sensitivity to low mass objects or good enough  velocity resolution to identify the majority of dark galaxies. 
\item If dark galaxies cluster like galaxies with stars then it is easy to confuse a dark galaxy detection with an incorrect optical source in the same neighbourhood.
\end{enumerate}

 Future surveys using the Arecibo multi-beam instrument ALFA should reveal many hundreds of dark galaxies - if they exist.

\end{document}